%% file: growing_cyl_arxiv.tex
\documentclass[a4paper]{amsart}
\usepackage[T1]{fontenc}
\usepackage[utf8]{inputenc}
\usepackage[italian, english]{babel}

\usepackage{amsmath}
\usepackage{amssymb}
\usepackage{amsthm}
\usepackage{hyperref}
\usepackage{subfig}
\usepackage{graphicx}
\usepackage{amsaddr}
\usepackage{xcolor}
\hypersetup{
    colorlinks,
    linkcolor={red!50!black},
    citecolor={blue!50!black},
    urlcolor={blue!80!black}
}
\author{D.~Riccobelli and P.~Ciarletta}
\address{MOX -- Dipartimento di Matematica, Politecnico di Milano,\\
piazza Leonardo da Vinci 32, 20133 Milano, Italy.}
\email{davide.riccobelli@polimi.it, pasquale.ciarletta@polimi.it}
\input{defs}
\begin{document}
\title{Morpho--elastic model of the tortuous tumour vessels}

\begin{abstract}
Solid tumours have the ability to assemble their own vascular network for optimizing their access to the vital nutrients. These new capillaries are morphologically different from normal physiological vessels. In particular, they have a much higher spatial tortuosity forcing an impaired flow within the peritumoral area. This is a major obstacle for the efficient delivery of antitumoral drugs.
This work proposes a morpho--elastic model of the tumour vessels. A tumour capillary is considered as a growing hyperelastic tube that is spatially constrained by a linear elastic environment, representing the interstitial matter. We assume that the capillary is an incompressible neo--Hookean material, whose growth is modeled using a multiplicative decomposition of the deformation gradient.
We study the morphological stability of the capillary by means of the method of incremental deformations superposed on finite strains, solving the corresponding incremental problem using the Stroh formulation and the impedance matrix method.  The incompatible axial growth of the straight capillary is found to control the onset of a bifurcation towards a tortuous shape. The post-buckling morphology is studied using a mixed finite element formulation in the fully nonlinear regime. The proposed model highlights how the geometrical and the elastic properties of the capillary and the surrounding medium concur to trigger the loss of marginal stability of the straight capillary and the nonlinear development of its spatial tortuosity. 
\end{abstract}

\maketitle

\section{Introduction}

Living matter has the ability to change its macroscopic shape even in absence of external forces, thanks to the activation of microscopic rearrangement processes such as growth and remodelling \cite{ambrosi2011perspectives}. In fact, whenever such underlying transformations introduce a geometrical incompatibility in the micro--structure, a state of internal stress arises in the material in order to accommodate these misfits. The accumulation of such internal stresses beyond a critical threshold may drive the onset of an elastic bifurcation. Morpho--elastic models successfully describe many  morphological transitions in living and inert matter  \cite{amar2005growth, goriely2008nonlinear, de2011nonlinear, destrade2017wrinkles}.

Recently, several works have addressed the problem of the stability of cylindrical structures subjected to differential growth and geometrical constraints. In \cite{moulton2011circumferential}, Moulton and Goriely studied the buckling of an hollow cylindrical tube subjected to a radial and circumferential differential growth. Subsequently O'Keeffe et al. \cite{o2013growth} addressed the problem of the stability of a solid cylinder growing along the axial direction, embedded into an elastic inert matrix and confined between two parallel, rigid planes. The stability of residually stressed cylindrical structures has been further studied by exploiting an alternative approach, prescribing the residual stress field instead of the growth tensor \cite{merodio2013influence, ciarletta2016morphology}.

 This work aims at modelling the morphogenesis of the tumour vascular network.  After an initial avascular phase, a solid tumour can activate a process known as angiogenesis, assembling its own vascular network for opening a new access to the vital nutrients \cite{levine2001mathematical, ALARCON2005369}. These new capillaries are morphologically different from normal physiological vessels. In particular, they have a much higher spatial tortuosity and an increased permeability \cite{stoll2003quantitative, bullitt2005vessel} forcing an impaired flow within the peritumoral area.  These structural peculiarities represent a major obstacle for the efficient delivery of antitumoral drugs \cite{jain1994barriers}.
 
In \cite{araujo2004new} Araujo and McElwain have proposed a model of the growth induced residual stress in solid tumors, assuming that the buckling of capillaries is induced by the stress applied by the tumor on the vessel. A seminal morpho--elastic model of this biological process has proved that an incompatible growth process of the tumor intertium can explain the buckling of capillaries \cite{maclaurin2012buckling} but not as easily their tortuosity. The aim of this work is to study the stability of a growing hyperelastic hollow cylinder taking into account for the linear elastic constraint of the surrounding interstitial matter. Contrarily to the work of MacLaurin et al. \cite{maclaurin2012buckling}, we assume that the buckling of the tumor capillary is not triggered by the growth of the surrounding tissue growth but by the growth of the vessel wall. 

This article is organized as follows.  In Section \ref{sec:elastic_model}, we introduce the morpho--elastic model and we derive the basic axis-symmetric solution of the corresponding hyperelastic problem.

In Section \ref{sec:lin_stab}, we perform a linear stability analysis of the basic axis-symmetric solution using the the method of incremental deformations superposed on a finite strain. 

In Section \ref{sec:numerical}, we describe the mixed finite element method that we have implemented to perform the numerical simulations of the post-buckling behavior.  The results of both the theoretical analysis and the numerical simulations are finally discussed in Section \ref{sec:disc}, together with some concluding remarks.

\section{The elastic model}
\label{sec:elastic_model}

Let the reference configuration of the elastic body be the open set $\Omega_0\subset\R^3$ such that
\[
\Omega_0 = \left\{\vect{X} = \left(R\cos\Theta,\,R\sin\Theta,\,Z)\;|\;R_\text{i}<R<R_\text{o}\text{ and }0<Z<H\right)\right\},
\]
representing the wall of the tumor capillary, composed by the endothelium and the basement membrane \cite{fung2013biomechanics}, where $R$, $\Theta$ and $Z$ are the cylindrical coordinates of the material point $\vect{X}$. We denote by $\vect{E}_R,\,\vect{E}_\Theta$ and $\vect{E}_Z$  the orthonormal vector basis in a cylindrical reference system.

We indicate by $\Omega$ the deformed configuration of the elastic tube and the mapping by 
\[
\vect{\varphi}:\Omega_0\rightarrow\Omega
\]
such that $\vect{u}(\vect{X})=\vect{\varphi}(\vect{X})-\vect{X}$ is the displacement vector and  $\tens{F} = \Grad\vect{\varphi}= \frac{\partial \vect{\varphi}}{\partial \vect{X}}$ be the deformation gradient.

The volumetric  growth of the body is enforced by introducing a multiplicative  decomposition of the deformation gradient \cite{kroner1959allgemeine, Lee1969,rodriguez1994stress},  as follows
\[
\tens{F} = \tens{F}_\text{e}\tens{G}
\]
so that $\tens{G}$ describes the metric distortion induced by the growth and $\tens{F}_\text{e}$ is the elastic deformation of the material restoring the geometrical compatibility of the current configuration.

We assume that the material is hyperelastic and incompressible, since the tissue constituents are mostly made of water. Denoting by $\psi$ its strain energy density per unit volume, the first Piola--Kirchhoff and the Cauchy stress tensors read
\begin{equation}
\label{eq:stress_tens}
\left\{
\begin{aligned}
&\tens{P}=\det\tens{G}\frac{\partial\psi(\tens{F}\tens{G}^{-1})}{\partial\tens{F}}-p\tens{F}^{-1}\\
&\tens{T} = \frac{1}{\det\tens{F}}\tens{F}\tens{P}
\end{aligned}
\right.
\end{equation}
where $p$ is the Lagrangian multiplier that enforces the incompressibility constraint $\det\tens{F}_\text{e}=1$.

Assuming quasi-static conditions in absence of external body forces, the balance of the linear and of the angular momentum reads
\begin{equation}
\label{eq:bilanMomen}
\Diver \tens{P} = \vect{0}\text{ in }\Omega_0\quad\text{or}\quad\diver\tens{T} = \vect{0}\text{ in }\Omega
\end{equation}
where $\Diver$ and $\diver$ denote the divergence operator in material and current coordinates, respectively.

The nonlinear system of equations \eqref{eq:bilanMomen} is complemented by the following boundary conditions
\begin{equation}
\label{eq:BCS}
\left\{
\begin{aligned}
&\tens{P}^T\vect{N} = \vect{0} &&\text{for }R = R_\text{i}\\
&\tens{P}^T\vect{N} = -\mu_k\vect{u} &&\text{for }R = R_\text{o}\\
&\tens{P}^T\vect{N}\cdot\vect{E}_R = 0 &&\text{for }Z = 0,\,H\\
&\tens{P}^T\vect{N}\cdot\vect{E}_\Theta = 0 &&\text{for }Z = 0,\,H\\
&u_Z = 0 &&\text{for }Z = 0,\,H\\
\end{aligned}
\right.
\end{equation}
where $\vect{N}$ denotes the outer normal in the Lagrangian configuration and $\mu_k$ is the linear elastic stiffness of the outer peritumoral tissue. Since the intercapillary distance is much bigger than the characteristic diameter of the capillary, we indeed assume that the outer tissue exerts a linear elastic response that is simplified by an isotropic spring foundation.

\subsection{Constitutive assumptions and basic axis-symmetric solution}

We assume that the tube is composed of an incompressible neo--Hookean material, thus the strain energy $\psi$ is given by
\begin{equation}
\label{eq:energy}
\psi(\tens{F})=\frac{\mu}{2}\left(I_1-3\right) = \frac{\mu}{2}\left(\lambda_1^2+\lambda_2^2+\lambda_3^2-3\right)
\end{equation}
where $I_1$ is the trace of the right Cauchy--Green tensor $\tens{C} = \tens{F}^T\tens{F}$ and $\lambda_i$ are the eigenvalues of the deformation gradient. We can write the Cauchy stress tensor \eqref{eq:stress_tens} as
\begin{equation}
\label{eq:CauchyneoHook}
\tens{T} = \mu\tens{F}\tens{G}^{-1}\tens{G}^{-T}\tens{F}^T-p\tens{I}
\end{equation}
where $\tens{I}$ is the identity tensor. We further assume that the growth tensor $\tens{G}$ has the form
\begin{equation}
\label{eq:growthtens}
\tens{G} = \diag(1,\,1,\,\gamma),
\end{equation}
so that the elastic tube grows along the axial direction. We look for a solution of the form
\[
\vect{\varphi}(\vect{X}) = r(R)\vect{E}_R+Z \vect{E}_Z.
\]
We denote by $r_\text{i} = r(R_\text{i})$ and $r_\text{o} = r(R_\text{o})$. For the sake of simplicity, in the following we omit the explicit dependence of $r$ on the variable $R$. The deformation gradient is given by
\begin{equation}
\label{eq:baseF}
\tens{F} = \diag\left(r',\,\frac{r}{R},\,1\right)
\end{equation}

Considering the equations \eqref{eq:growthtens} and \eqref{eq:baseF}, the incompressibility constraint $\det \tens{F}_\text{e}=1$ leads to the following differential equation
\begin{equation}
\label{eq:incomr}
r'r = \gamma R
\end{equation}
so that
\begin{equation}
\label{eq:r}
r = \sqrt{\gamma (R^2 - R_\text{i}^2)+ r_\text{i}^2}.
\end{equation}
By enforcing the global incompressibility constraint in Eq.\eqref{eq:r}, we get
\begin{equation}
\label{eq:ro}
r_\text{o}=\sqrt{\gamma (R_\text{o}^2 - R_\text{i}^2)+ r_\text{i}^2}.
\end{equation}

The inverse of  Eq. \eqref{eq:r} reads:
\[
R = \sqrt{\frac{r^2 - r_\text{i}^2}{\gamma}+ R_\text{i}^2},
\]
so that, from \eqref{eq:incomr}, we get:
\begin{equation}
\label{eq:rprimo}
r' = \frac{\gamma R}{r} = \frac{\sqrt{\gamma\left(r^2 - r_\text{i}^2\right)+ \gamma^2 R_\text{i}^2}}{r}.
\end{equation}

From \eqref{eq:CauchyneoHook}, \eqref{eq:growthtens} and \eqref{eq:baseF} the Cauchy stress tensor reads
\[
\tens{T} = T_{rr}\vect{e}_r\otimes\vect{e}_r+T_{\theta\theta}\vect{e}_\theta\otimes\vect{e}_\theta+T_{zz}\vect{e}_z\otimes\vect{e}_z
\]
where $\vect{e}_r$, $\vect{e}_\theta$ and $\vect{e}_z$ constitute the local orthonormal vector basis of the actual configuration in cylidrical coordinates and
\[
\left\{
\begin{aligned}
&T_{rr}(r) = \mu r'^2-p,\\
&T_{\theta\theta}(r) = \mu\frac{r^2}{R^2}-p,\\
&T_{zz}(r) = \mu\frac{1}{\gamma^2}-p.\\
\end{aligned}
\right.
\]

In cylindrical coordinates, the balance of the linear and angular momentum \eqref{eq:bilanMomen} reads
\begin{equation}
\label{eq:balancemomT}
\frac{dT_{rr}}{dr}+\frac{T_{rr}-T_{\theta\theta}}{r}=0;
\end{equation}
with the following boundary conditions \eqref{eq:BCS}:
\begin{equation}
\label{eq:Tbcs}
\left\{
\begin{aligned}
&T_{rr}(0) = 0 &&\text{for }r = r_\text{i}\\
&T_{rr}(r_\text{o}) = -\mu_k\frac{R_\text{o}}{r_\text{o}}(r_\text{o}-R_\text{o}) &&\text{for }r = r_\text{o}\\
\end{aligned}
\right.
\end{equation}

Making use of \eqref{eq:Tbcs}, we can integrate the equation \eqref{eq:balancemomT} from $r=r_\text{i}$ to $r=r_\text{o}$, obtaining
\[
\mu_k\frac{R_\text{o}}{r_\text{o}}(r_\text{o}-R_\text{o}) = \int_{r_\text{i}}^{r_\text{o}}\frac{T_{rr}(r)-T_{\theta\theta}(r)}{r}dr,
\]
so that, together with the equation \eqref{eq:ro}, we obtain an equation for $r_\text{i}$ which can be solved numerically if we fix the ratio $R_\text{o}/R_\text{i}$ and the axial growth parameter $\gamma$.

Finally, we integrate the equation \eqref{eq:balancemomT} from $r_\text{i}$ to $r$ in order  to determine the Lagrangian multiplier $p$, so that
\[
p = \mu r'^2 + \int_{r_\text{i}}^{r}\frac{T_{rr}(s)-T_{\theta\theta}(s)}{s}ds.
\]
The latter integral can be computed analytically, obtaining 
\begin{equation}
\label{eq:pres_base}
\begin{aligned}
p(r) = 
\frac{1}{2} \gamma  \mu  &\left(\frac{\gamma  R_\text{i}^2-r_\text{i}^2}{r^2}-\log \left(r^2-r_\text{i}^2+\gamma  R_\text{i}^2\right)+2 \log (r)+\right.\\
&\left.+\frac{\gamma  R_\text{i}^2}{r_\text{i}^2}-2 \log (r_\text{i})+\log \left(\gamma  R_\text{i}^2\right)+1\right).
\end{aligned}
\end{equation}

Thus, we have found a basic axis-symmetric solution of the boundary value problem, that is given by Eqs.(\ref{eq:r},\ref{eq:pres_base})

In the following, we study its marginal stability  as a function of the control parameter $\gamma$ denoting the local volumetric growth along the axial direction.

\section{Linear stability analysis}
\label{sec:lin_stab}

In this section we study the linear stability of the finitely deformed tube by using the method of incremental deformations superposed on a finite strain \cite{ogden1997non}.

We rewrite the resulting incremental boundary value problem into a more convenient form called Stroh formulation and we implement a numerical method based on the impedance matrix method to solve it.

\subsection{Incremental boundary value problem}

We denote the incremental displacement field $\delta\vect{u}$. Let $\tens{\Gamma}= \grad\delta\vect{u}$, we introduce the push-forward of the  incremental Piola--Kirchhoff stress in the finitely deformed configuration of the axis-symmetric solution, that is given by
\begin{equation}
\label{eq:incCon}
\delta\tens{P}_0=\mathcal{A}_0:\tens{\Gamma}+p\tens{\Gamma}-\delta p \tens{I},\qquad\text{where}\qquad (\mathcal{A}_0:\tens{\Gamma})_{ij}=A_{0ijhk}\Gamma_{kh},
\end{equation}
where $\mathcal{A}_0$ is the fourth order tensor of instantaneous elastic moduli, $\delta p$ is the increment of the Lagrangian multiplier that imposes the incompressibility constraint, and the convention of summation over repeated indices is adopted.

The components of the tensor $\mathcal{A}_0$ for a neo--Hookean material, are given by
\[
A_{0ijhk}=\mu\delta_{jk}(B_\text{e})_{ih} =\mu\delta_{jk}\delta_{ih}(\lambda_i^\text{e})^2 
\]
where $\tens{B}_\text{e}=\tens{F}_\text{e}\tens{F}_\text{e}^T$, $\lambda_i^\text{e}$ are the eigenvalues of the tensor $\tens{F}_\text{e}=\tens{F}\tens{G}^{-1}$. Considering the growth tensor \eqref{eq:growthtens},the deformation gradient \eqref{eq:baseF} and the equation \eqref{eq:rprimo}, such eigenvalues are given by
\[
\left\{
\begin{aligned}
&\lambda_1^\text{e}=r' = \frac{\sqrt{\gamma\left(r^2 - r_\text{i}^2\right)+ \gamma^2 R_\text{i}^2}}{r},\\
&\lambda_2^\text{e}=\frac{r}{R} = \frac{\gamma r}{\sqrt{\gamma\left(r^2 - r_\text{i}^2\right)+ \gamma^2 R_\text{i}^2}},\\
&\lambda_3^\text{e} = \frac{1}{\gamma}.
\end{aligned}
\right.
\]

The incremental form of the balance of the linear momentum and of the incompressibility constraint are given by
\begin{equation}
\label{eq:incPiola}
\left\{
\begin{aligned}
&\diver \delta\tens{P}_0=\vect{0}, &&\text{in }\Omega,\\
&\tr\tens{\Gamma}=0 &&\text{in }\Omega.
\end{aligned}
\right.
\end{equation}

This system of partial differential equations is complemented by the following boundary conditions
\begin{equation}
\label{eq:incBCS}
\left\{
\begin{aligned}
&\delta\tens{P}^T\vect{e}_r = \vect{0} &&\text{for }r = r_\text{i}\\
&\delta\tens{P}^T\vect{e}_r = -\mu_k\frac{R_\text{o}}{r_\text{o}}\delta\vect{u} &&\text{for }r = r_\text{o}\\
&\delta\tens{P}\vect{e}_r\cdot\vect{e}_r = 0 &&\text{for }z = 0,\,H\\
&\delta\tens{P}^T\vect{e}_r\cdot\vect{e}_\theta = 0 &&\text{for }z = 0,\,H\\
&\delta u_z = 0 &&\text{for }z = 0,\,H.\\
\end{aligned}
\right.
\end{equation}
where $\vect{e}_r,\,\vect{e}_\theta$ and $\vect{e}_z$ is the vector basis in cylindrical coordinates in the actual configuration.

To implement a robust numerical method, we employ a method which is different to the one used in \cite{o2013growth} where the authors studied the stability of a growing solid cylinder surrounded by an elastic tube. We reformulate the boundary value problem given by the equations~\eqref{eq:incPiola}--\eqref{eq:incBCS} by using the Stroh formulation.

\subsection{Stroh formulation}

We denote with $u,\,v$ and $w$ the components of $\delta\vect{u}$ in cylindrical coordinates.%, the tensor $\tens{\Gamma}$ is given by
%\begin{equation}
%\tens{\Gamma}=\left(
%\begin{aligned}
% &\partial_r u && \frac{\partial_\theta u-v}{r} && \partial_z u \\
% &\partial_r v && \frac{u+\partial_\theta v}{r} && \partial_z v \\
% &\partial_r w && \frac{\partial_\theta w}{r} && \partial_z w \\
%\end{aligned}
%\right)
%\end{equation}
To reduce the system of partial differential equations \eqref{eq:incPiola} to a system of ordinary differential equations, we assume the following ansatz \cite{maclaurin2012buckling}:
\begin{gather*}
u(r,\,\theta,\,z) = U(r)\cos(m\theta)\cos(k z),\\
v(r,\,\theta,\,z) = V(r)\sin(m\theta)\cos(k z),\\
w(r,\,\theta,\,z) = W(r)\cos(m\theta)\sin(k z),\\
\end{gather*}
where $m\in\mathbb{N}$ and $k\in\R$ with $k\geq0$.

Following the procedure exposed in \cite{balbi2013morpho}, we consider the components $\delta P_{rr},\,\delta P_{\theta r}$ and $\delta P{z r}$ as additional unknowns. We assume then that
\begin{gather}
\label{eq:prr}
\delta P_{rr}(r,\,\theta,\,z) = p_{rr}(r)\cos(m\theta)\cos(k z),\\
\delta P_{r\theta}(r,\,\theta,\,z) = p_{r\theta}(r)\sin(m\theta)\cos(k z),\\
\delta P_{rz}(r,\,\theta,\,z) = p_{rz}(r)\cos(m\theta)\sin(k z).
\end{gather}

We substitute \eqref{eq:prr} into \eqref{eq:incCon} obtaining the following expression for $\delta p$:
\begin{footnotesize}
\[
\begin{aligned}
\delta p &= \frac{\gamma  \mu  U'(r) \left(\gamma  R_\text{i}^2 \left(r^2+3 r_\text{i}^2\right)+r^2 r_\text{i}^2 \left(-\log \left(r^2-r_\text{i}^2+\gamma  R_\text{i}^2\right)+2 \log (r)-2 \log (r_\text{i})+\log \left(\gamma  R_\text{i}^2\right)\right)\right)}{2 r^2 r_\text{i}^2}\\
&\qquad+\frac{3\gamma  \mu  U'(r) (r-r_\text{i}) (r+r_\text{i})}{2 r^2}-p_{rr}(r)
\end{aligned}
\]
\end{footnotesize}
We introduce the displacement-traction vector $\vect{\eta}$ as
\begin{equation}
\label{eq:eta}
\vect{\eta} = \left[\vect{U},\,r\vect{T}\right]\quad\text{ where }\left\{
\begin{aligned}
&\vect{U} = \left[U,\,V,\,W\right],\\
&\vect{T} = \left[p_{rr},\,p_{r\theta},\,p_{rz}\right].
\end{aligned}
\right.
\end{equation}

By using a well-established procedure \cite{stroh1962steady}, exploiting the incremental constitutive relations \eqref{eq:incCon}, we can rewrite the incremental system of partial differential equations \eqref{eq:incPiola} as
\begin{equation}
\label{eq:Stroh}
\frac{d\vect{\eta}}{dr}=\frac{1}{r}\tens{N}\vect{\eta}
\end{equation}
where $\tens{N}\in\R^{6\times6}$ is the Stroh matrix; the expressions of its components are reported in the appendix. In particular, we can identify four sub-blocks
\[
\tens{N}=\begin{bmatrix}
\tens{N}_1 &\tens{N}_2\\
\tens{N}_3 &\tens{N}_4
\end{bmatrix}
\]
such that $\tens{N}_i\in\R^{3\times3}$ and $
\tens{N}_1 = -\tens{N}_4^T,\quad \tens{N}_2=\tens{N}_2^T,\quad\tens{N}_3=\tens{N}_3^T$.

\subsection{Impedance matrix method}
The system of ordinary differential equations gien by Eq. \eqref{eq:Stroh} is numerically solved using the impedance matrix method \cite{biryukov1985impedance, biryukov1995surface}.
We introduce the matricant
\[
\tens{M}(r,\,r_\text{i}) = \begin{bmatrix}
\tens{M}_1(r,\,r_\text{i}) &\tens{M}_2(r,\,r_\text{i})\\
\tens{M}_3(r,\,r_\text{i}) &\tens{M}_4(r,\,r_\text{i})
\end{bmatrix},\qquad\tens{M}\in\R^{6\times6}
\]
called conditional matrix. Such a matrix is a solution of the problem
\begin{equation}
\label{eq:defM}
\left\{
\begin{aligned}
&\frac{d}{dr}\tens{M}(r,\,r_\text{i})=\frac{1}{r}\tens{N}\tens{M}(r,\,r_\text{i}),\\
&\tens{M}(r_\text{i},\,r_\text{i})=\tens{I}.
\end{aligned}
\right.
\end{equation}

It is easy to verify that the solution of the Stroh equation \eqref{eq:Stroh} is given by
\begin{equation}
\label{eq:etaM}
\vect{\eta}(r)=\tens{M}(r,\,r_\text{i})\vect{\eta}(r_\text{i}).
\end{equation}

Since $\vect{T}(r_\text{i})=\vect{0}$, exploiting the relation \eqref{eq:etaM}, we can define the conditional impedance matrix $\tens{Z}(r,\,r_\text{i})$ \cite{norris2010wave} as
\begin{equation}
\label{eq:Z}
\tens{Z}(r,\,r_\text{i}) = \tens{M}_3(r,\,r_\text{i})\tens{M}_1^{-1}(r,\,r_\text{i}).
\end{equation}

For the sake of simplicity we omit the explicit dependence of $\tens{Z}$ on $r$ and $r_\text{i}$. Such a matrix satisfy the following relation
\[
r\vect{T}=\tens{Z}\vect{U}\qquad \forall r\in(r_\text{i},\,r_\text{o}).
\]

Thus, we can observe that the Stroh system \eqref{eq:Stroh} can be written as
\begin{gather}
\label{eqn:U'}
\frac{d\vect{U}}{d r}=\frac{1}{r}(\tens{N}_1+\tens{N}_2\tens{Z})\vect{U},\\
\label{eqn:ZU'}
\frac{d\tens{Z}}{dr}\vect{U}+\tens{Z}\frac{d\vect{U}}{dr}=\frac{1}{r}(\tens{N}_3+\tens{N}_4\tens{Z})\vect{U}.
\end{gather}

We now can substitute \eqref{eqn:U'} in \eqref{eqn:ZU'} obtaining a Riccati differential equation
\begin{equation}
\label{eq:Riccati}
\frac{d\tens{Z}}{dr}=\frac{1}{r}(\tens{N}_3+\tens{N}_4\tens{Z}-\tens{Z}\tens{N}_1-\tens{Z}\tens{N}_2\tens{Z});
\end{equation}

As a starting condition, considering \eqref{eq:defM} and the definition of surface impedance matrix \eqref{eq:Z}, we set
\[
\tens{Z}(r_\text{i},\,r_\text{i})=\tens{0}.
\]

The boundary condition in \eqref{eq:incBCS} linked to the presence of the springs at $r = r_\text{o}$ can be written as
\[
\vect{T} = -\mu_k\frac{R_\text{o}}{r_\text{o}}\vect{U}
\]
so that
\[
\left(\tens{Z}+\mu_kR_o\tens{I}\right)\vect{U}=0
\]

Non-null solutions of the incremental problem exist if and only if
\begin{equation}
\label{eq:stop_cond}
\det\left(\tens{Z}+\mu_kR_o\tens{I}\right) = 0.
\end{equation} 

For a fixed value of the control parameter $\gamma$ we integrate the Riccati equation \eqref{eq:Riccati} from $r=r_\text{i}$ up to $r=r_\text{o}$ making use of the the software \textsc{Mathematica} (ver. 11.2, Wolfram Research, Champaign, IL, USA). We iteratively increase the control parameter $\gamma$  until the stop condition \eqref{eq:stop_cond} is satisfied.

\subsection{Marginal stability thresholds and critical modes}

\begin{figure}
\centering
\includegraphics[width=0.5\textwidth]{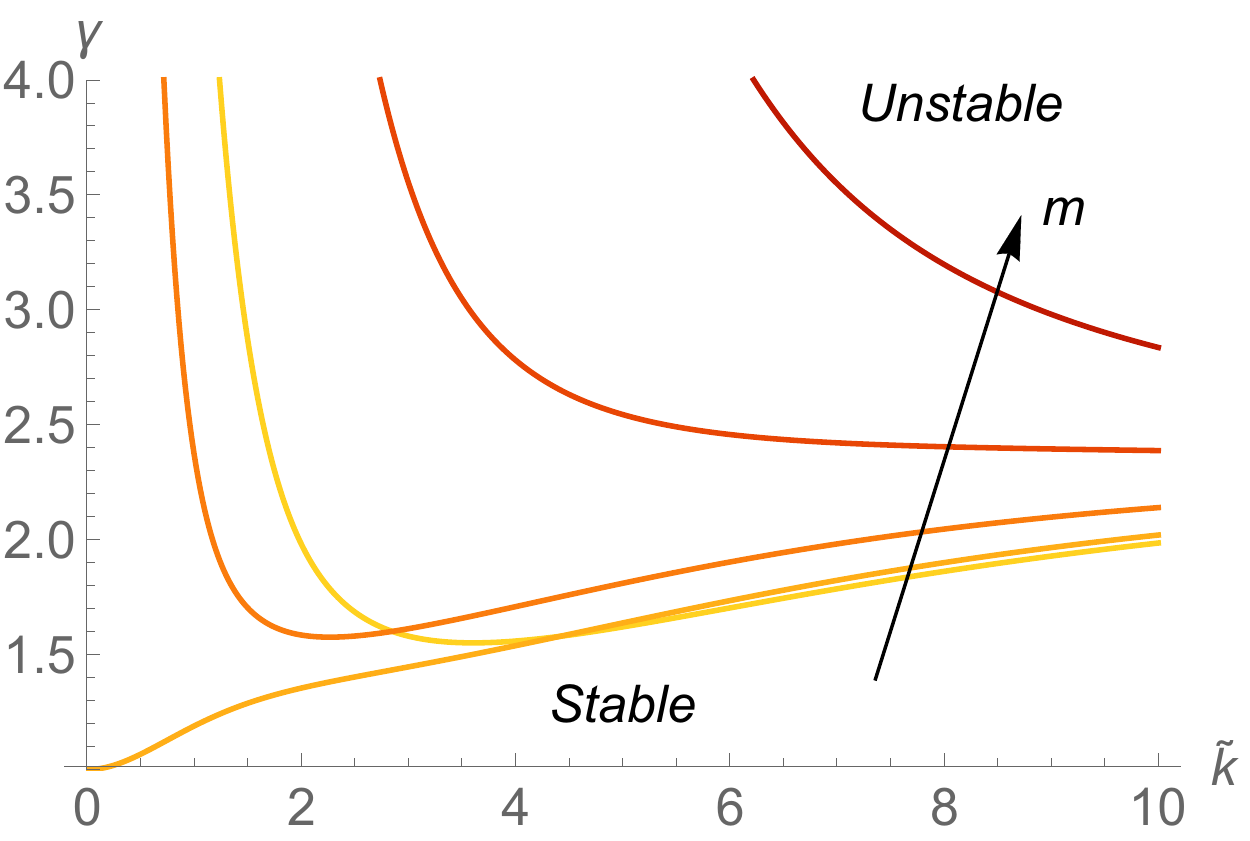}\includegraphics[width=0.5\textwidth]{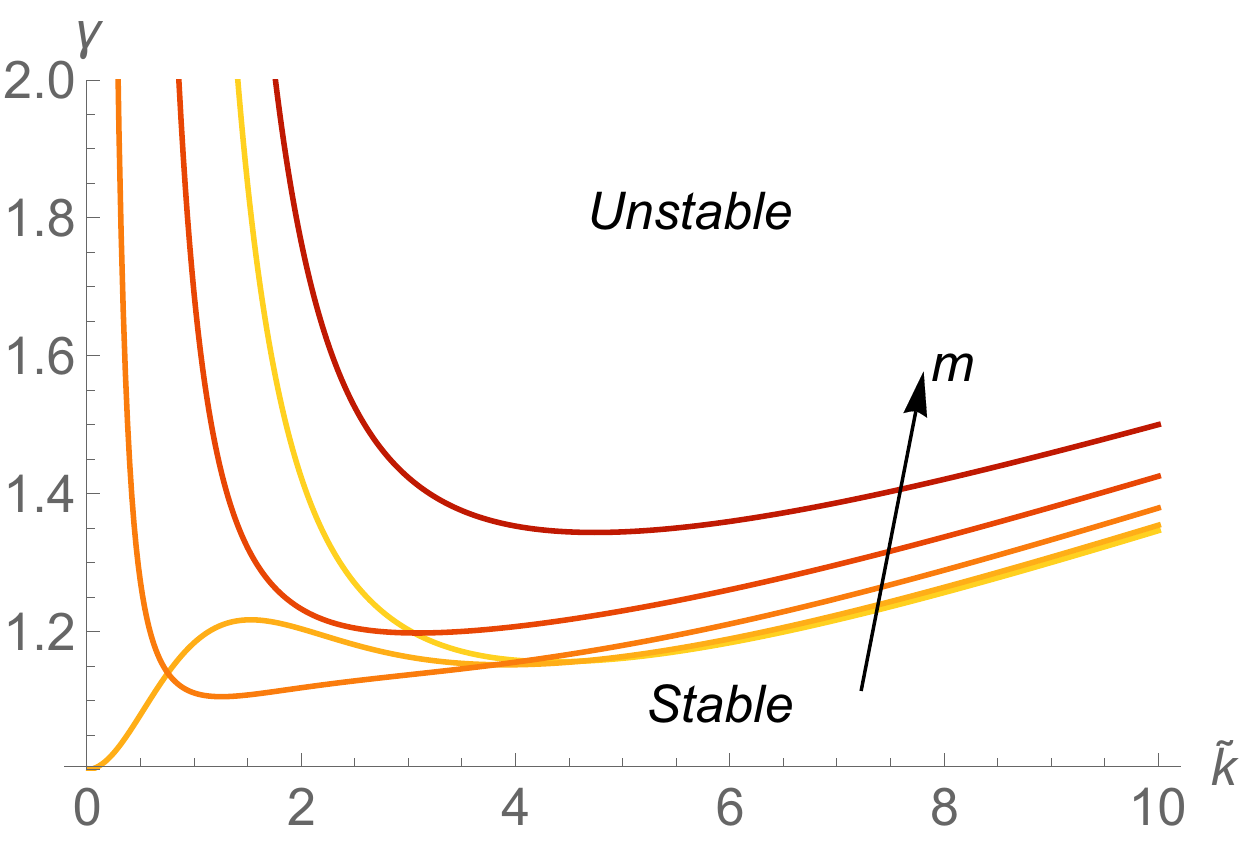}
\caption{Marginal stability curves $k R_\text{o}$ versus $\gamma$ when the elastic constant of the springs is $\mu_k=0$, the aspect ratio $\alpha_R$ is equal to $0.5$ (top) and $0.8$ (bottom). The circumferential wavenumber $m$ varies from $0$ (light line) up to $4$ (dark line), the arrow denotes the direction in which $m$ increases.}
\label{fig:marg_stab_kel_0_RiRo}
\end{figure}

In this section we discuss the results of the linear stability analysis.\\
\noindent Setting $\mu_k = 0$, we neglect the elastic contribution of the surrounding matter, thus dealing with a classical problem of Euler buckling. The corresponding  marginal stability curves are depicted in Fig.~\ref{fig:marg_stab_kel_0_RiRo}, in quantitative agreement with the results obtained by Goriely and co-workers \cite{goriely2008nonlinear}. As expected, the marginal stability threshold tends to $\gamma=1$ for $m=1$ and  $\tilde{k}$ tends to zero, i.e. the critical mode is the one with infinite wavelength along the axial direction.

the presence of an elastic foundation at the outer surface of the capillary  drastically changes this limiting behavior of Euler buckling. The elastic boundary value problem is governed by the following the dimensionless parameters:
\[
\tilde{k} = k R_\text{o},\qquad\alpha_k=\frac{\mu_k R_\text{o}}{\mu},\qquad\alpha_R = \frac{R_\text{i}}{R_\text{o}},
\]
where $\tilde{k}$ represents the dimensionless axial wavenumber, $\alpha_k$ is the ratio between the surface and bulk elastic energies,  and $\alpha_R$  is the geometrical aspect ratio of the tube. 

The radius of a tumour capillary measures $5.1\pm0.7\,\mu$m while its length $66.8 \pm 34.2 \,\mu$m \cite{less1991microvascular}. Thus, for a given length $L$, the admissible axial wavenumber $\tilde{k}$ are given by
\[
\tilde{k} = n \pi\frac{R_\text{o}}{L}\qquad n\in\mathbb{N}
\]
for the sake of simplicity, in the following we consider $\tilde{k}$ continuous since the slenderness ratio is small.

\begin{figure}
\centering
\includegraphics[width=0.5\textwidth]{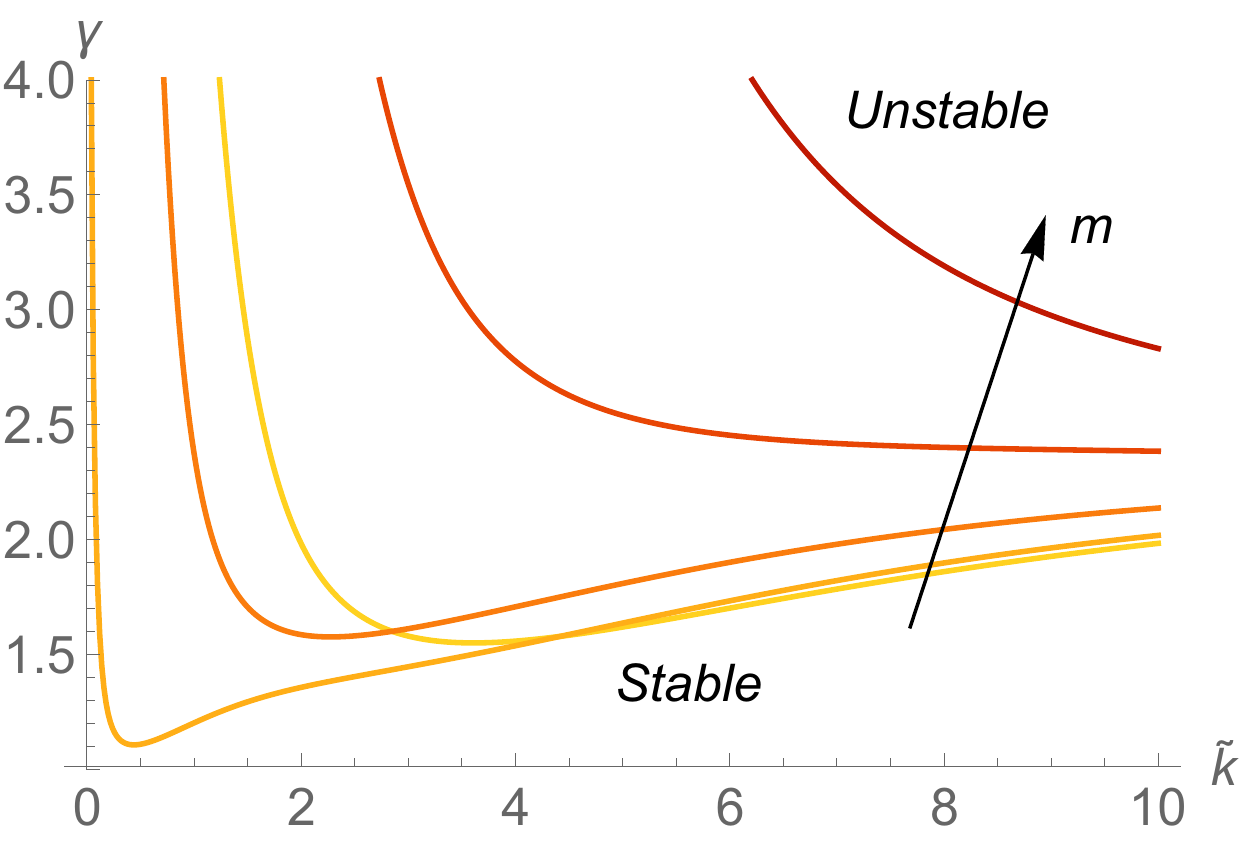}\includegraphics[width=0.5\textwidth]{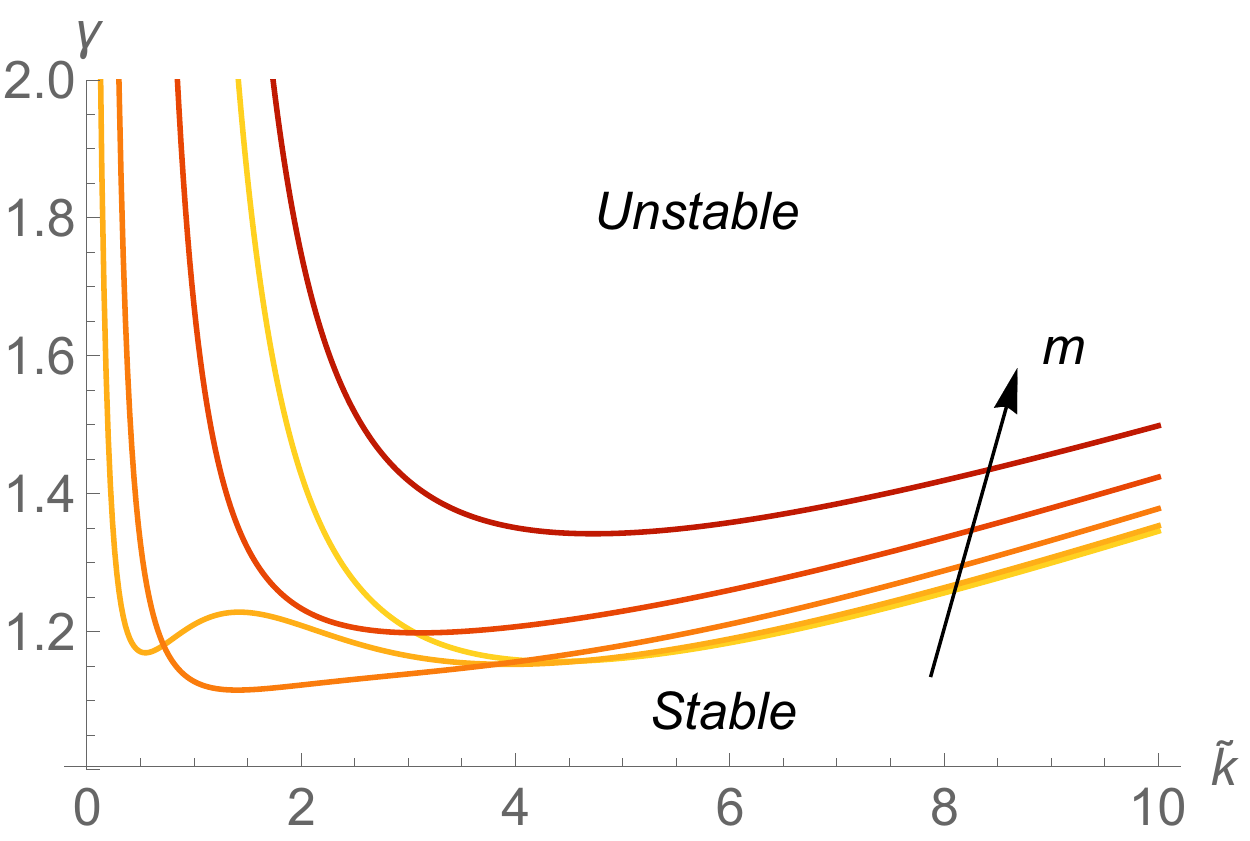}
\caption{Marginal stability curves $\tilde{k}$ versus $\gamma$ when $\alpha_k=0.01$, the aspect ratio $\alpha_R$ is equal to $0.5$ (top) and $0.8$ (bottom). The circumferential wavenumber $m$ varies from $0$ (light line) up to $4$ (dark line), the arrow denotes the direction in which $m$ increases.}
\label{fig:marg_stab_kel_001_RiRo}
\end{figure}

In Fig.~\ref{fig:marg_stab_kel_001_RiRo} we report the marginal stability curves when $\alpha_k = 0.01$. These marginal stability curves tend to the ones plotted in Fig.~\ref{fig:marg_stab_kel_0_RiRo} where $k R_\text{o}$ is large. However, a different behavior arises in the limit where $k R_\text{o}$ tends to zero, especially since the marginal stability threshold $\gamma$ now goes to infinity for $m=1$.

\begin{figure}
\centering
\includegraphics[width=0.5\textwidth]{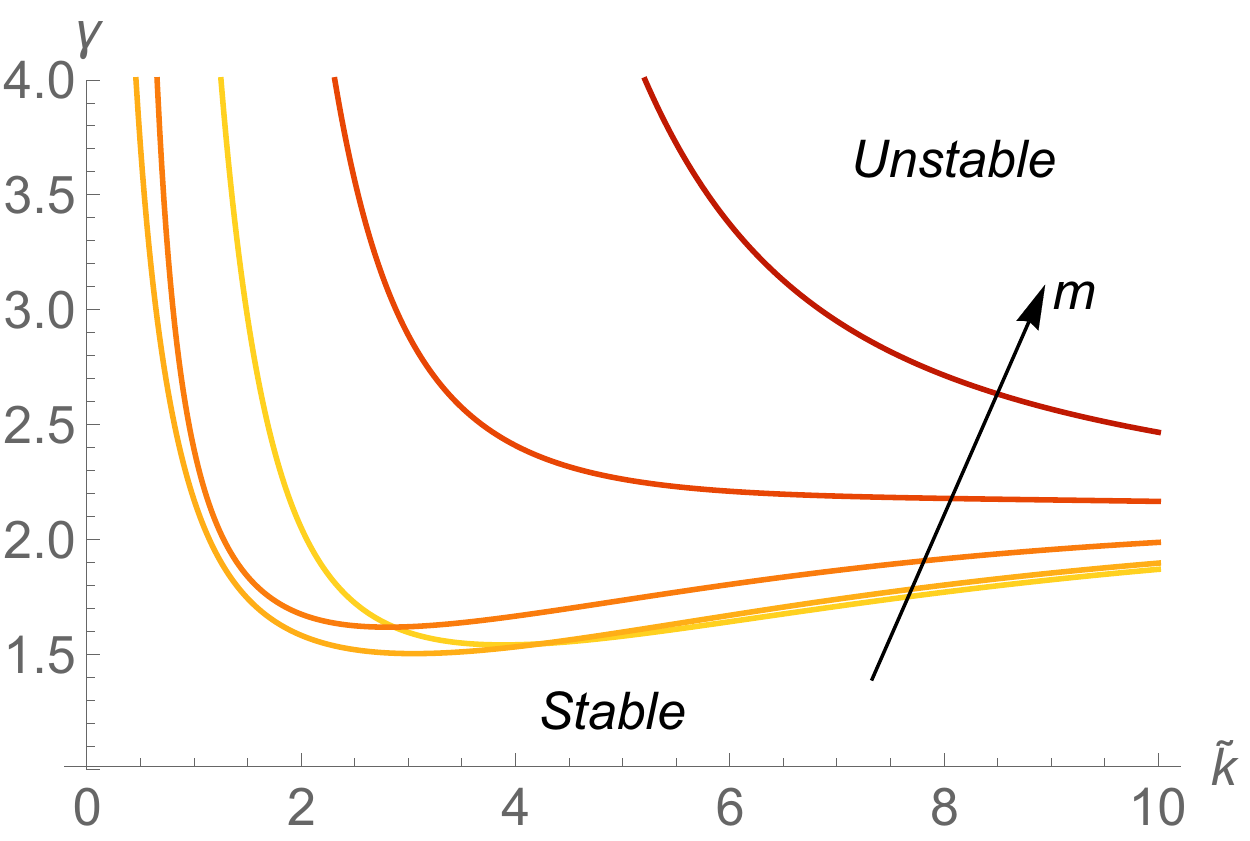}\includegraphics[width=0.5\textwidth]{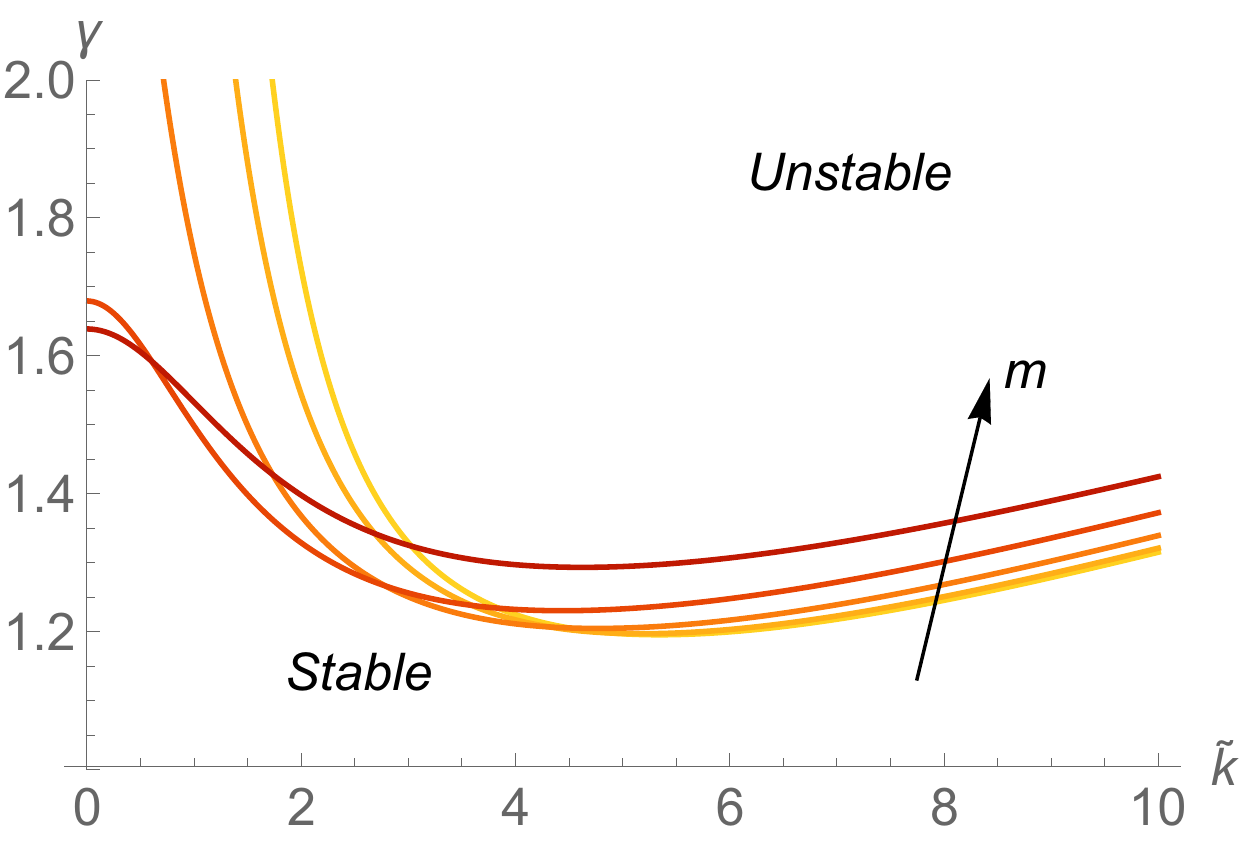}
\caption{Marginal stability curves $\tilde{k}$ versus $\gamma$ when $\alpha_k=1$, the aspect ratio $\alpha_R$ is equal to $0.5$ (top) and $0.8$ (bottom). The circumferential wavenumber $m$ varies from $0$ (light line) up to $4$ (dark line), the arrow denotes the direction in which $m$ increases.}
\label{fig:marg_stab_kel_1_RiRo}
\end{figure}

This effect  is even more evident by setting $\alpha_k=1$, as sketched in Fig.~\label{eq:marg_stab_kel_1_RiRo}.

For each fixed value of the dimensionless parameter $\alpha_k$, we define the critical value $\gamma_\text{cr}$ as the minimum value of the marginal stability curves $\gamma$ versus $\tilde{k}$ for all the circumferential wavenumber $m$. The corresponding axial and circumferential critical modes are denoted by $\tilde{k}_\text{cr}$ and $m_\text{cr}$, respectively.

\begin{figure}
\centering
\includegraphics[width=0.49\textwidth]{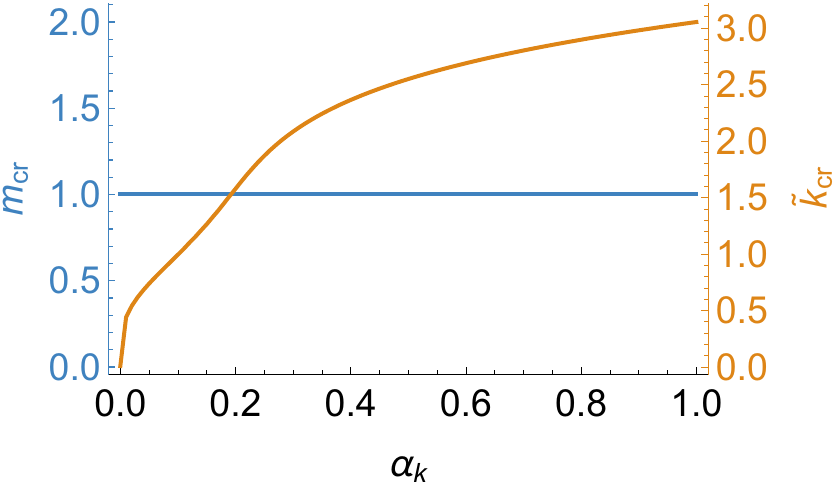}\;
\includegraphics[width=0.49\textwidth]{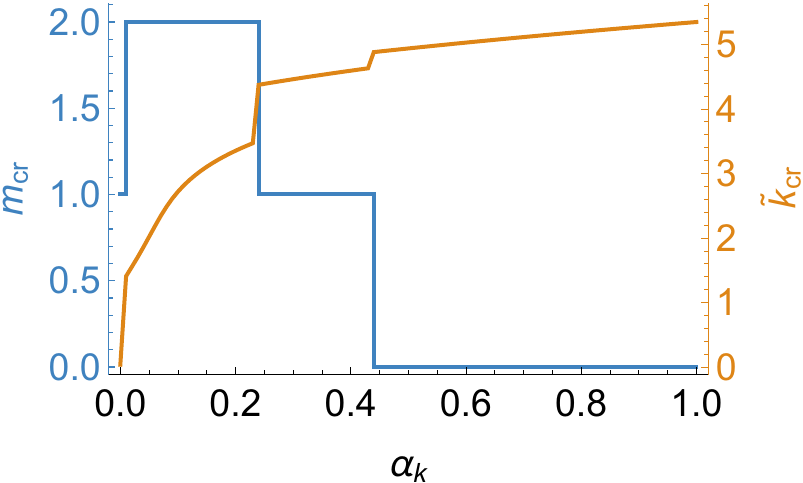}
\caption{Plots of the critical wavenumbers $m_\text{cr}$ and $\tilde{k}_\text{cr}$ versus $\alpha_k$ for $\alpha_R=0.5$ (left) and $\alpha_R=0.8$ (right).}
\label{fig:cr_mode}
\end{figure}

We plot the critical modes in Fig.~\ref{fig:cr_mode} for two different values of the aspect ratio, $\alpha_R=0.5$ (left)  and $0.8$ (right). In both cases the critical axial wavenumber $\tilde{k}_\text{cr}$ is increasing as $\alpha_k$ increases,  highlighting  discrete changes of the critical circumferential wavenumber. 

\begin{figure}
\centering
\includegraphics[width=0.45\textwidth]{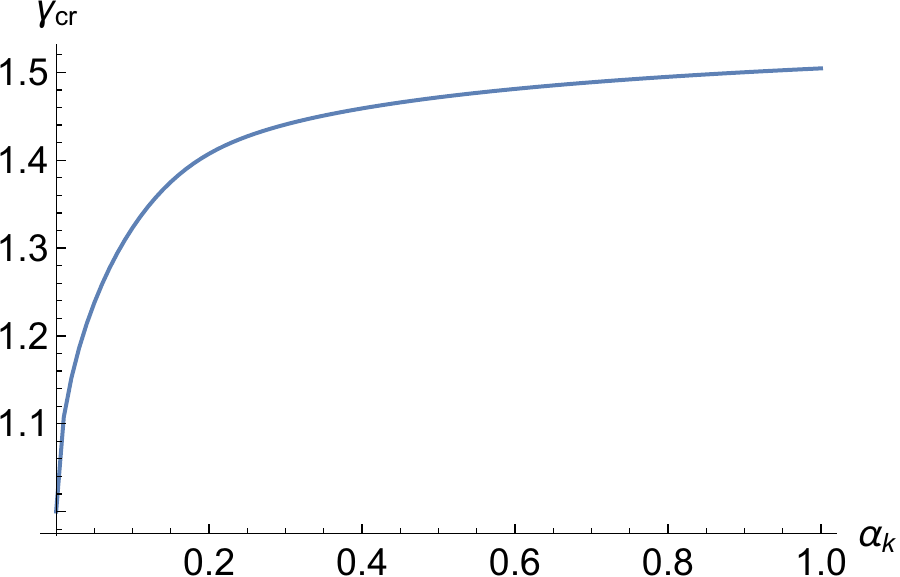}\;
\includegraphics[width=0.45\textwidth]{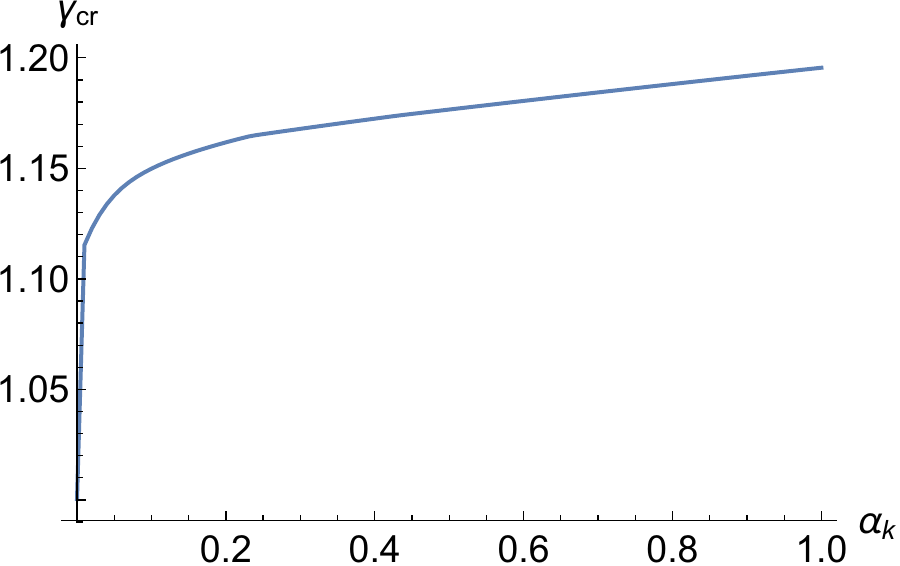}
\caption{Plots of the critical values of the control parameter $\gamma_\text{cr}$ versus $\alpha_k$ for $\alpha_R=0.5$ (left) and $\alpha_R=0.8$ (right).}
\label{fig:gammacr}
\end{figure}

Also the critical value of the control parameter $\gamma_\text{cr}$ is an increasing function of the dimensionless parameter $\alpha_k$ as shown in the plots of Fig.~\ref{fig:gammacr}.

We also compute the dimensionless critical load $\tau$ that is applied on the top surface in order to enforce the torsion through the application of a surface traction at the tube  top and bottom ends. Let  $\mathcal{S}=\{\vect{X}\in\Omega\;|\;Z=H\}$, this critical load is given by: 
\[
\tau = -\frac{1}{\mu  R_\text{o}^2}\int_\mathcal{S}P_{ZZ}dS = -\frac{1}{\mu R_\text{o}^2}\int_\mathcal{S}\left(\frac{\mu}{\gamma^2}-p\right)dS.
\]
where $\gamma_{cr}$ is the marginal stability threshold and $ L_{\text{cr}} = \pi R_\text{o}/\tilde{k}_\text{cr}$ is half of the critical wavelength.

\begin{figure}
\centering
\includegraphics[width=0.45\textwidth]{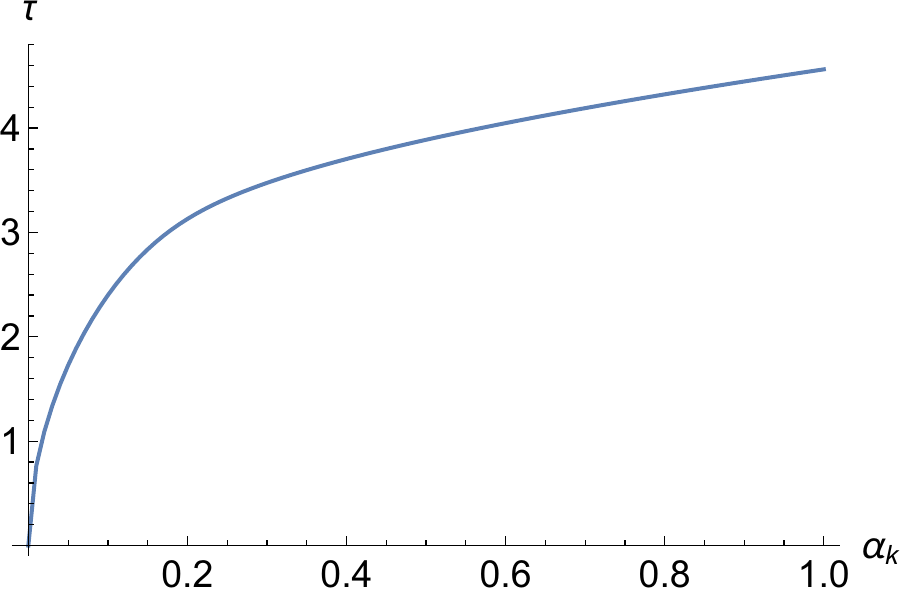}\;
\includegraphics[width=0.45\textwidth]{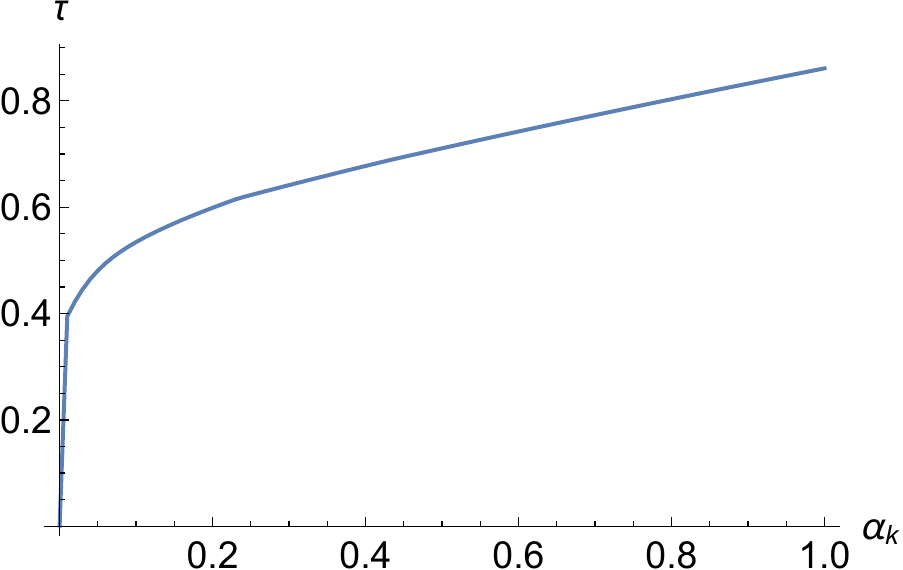}
\caption{Plots of the dimensionless critical load $\tau$ vs $\alpha_k$ for $\alpha_R=0.5$ (left) and $\alpha_R=0.8$ (right).}
\label{fig:tau}
\end{figure}

In Fig.~\ref{fig:tau} we plot $\tau$ versus $\alpha_k$ for $\alpha_R=0.5$ and $0.8$. In both cases, the critical load is a decreasing function of  $\alpha_k$,  so the presence of the outer elastic confinement has a stabilizing effect, whilst  thinner tubes always require lower critical loads.

\section{Post-buckling behaviour}
\label{sec:numerical}

In order to study the behavior of the buckled configuration far beyond the marginal stability threshold, we have implemented a finite element code to discretize and numerically solve the fully nonlinear boundary value problem given by \eqref{eq:bilanMomen}--\eqref{eq:BCS}.

\subsection{Finite-element implementation}

To break the axial symmetry of the problem, we numerically solve the  boundary value problem only on half cylinder whose height is half of the critical axial wavelength:
\[
\Omega_\text{c}=\left\{\vect{X}=(X,\,Y,\,Z)\;\bigg|\;\frac{R_\text{i}}{R_\text{o}}<\sqrt{X^2+Y^2}<1\;\cap\;0<Z<\frac{\pi}{\tilde{k}_\text{cr}}\;\cap\; Y>0\right\}.
\]
where $\tilde{k}_\text{cr}$ is the critical dimensionless wave-number arising from the linear stability analysis presented in Section \ref{sec:lin_stab} and $(\vect{E}_X,\,\vect{E}_Y,\,\vect{E}_Z)$ is the cartesian orthonormal vector basis.
We discretize this domain by using a tetrahedral mesh composed by 93398 elements. We used the Taylor--Hood $\vect{P}_2$--$P_1$ element, i.e. the displacement field is given by a continuous, piecewise quadratic function while the pressure field by a continuous, piecewise linear function. The choice of this particular element is motivated by its stability for non-linear elastic problems \cite{auricchio2013approximation}. Since we have only considered a half tube, we complement the boundary conditions \eqref{eq:BCS} by adding the following equations
\[
\left\{
\begin{aligned}
&\tens{P}^T\vect{E}_Y\cdot\vect{E}_X = 0 &&\text{for }Y = 0,\\
&\tens{P}^T\vect{E}_Y\cdot\vect{E}_Z = 0 &&\text{for }Y = 0,\\
&u_Y = 0 &&\text{for }Y = 0.\\
\end{aligned}
\right.
\]

\begin{figure}[t!]
\centering
\includegraphics[width=0.7\textwidth]{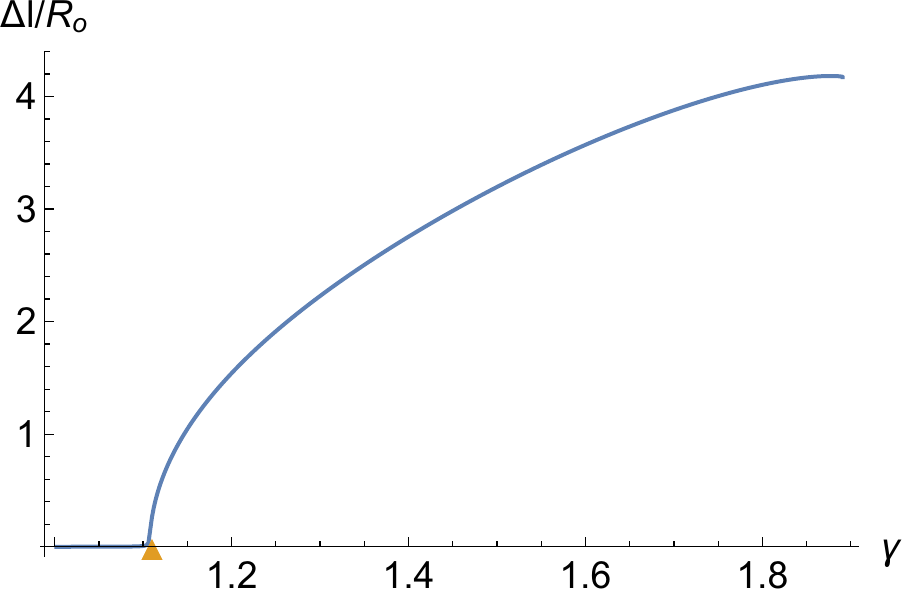}
\caption{Bifurcation diagram where we show the dimensionless  parameter $\Delta l/R_\text{o}$ versus the control parameter $\gamma$ when $\alpha_k=0.01$, $\alpha_R=0.5$. The numerical simulation is validated against  the marginal stability threshold computed with the linear stability analysis (orange triangle, $\gamma_\text{cr} = 1.1103$).}
\label{fig:bif}
\end{figure}

The numerical algorithm is based on a Newton continuation method \cite{seydel2009practical}, the control parameter $\gamma$ being incremented starting from $1$ with an automatic adaptation of step if the Newton method does not converge.

In order to follow the bifurcated branch, a small perturbation is imposed at the outer boundary of the cylinder according to the critical mode arising from the linear stability analysis \cite{budiansky1974theory, seydel2009practical}. The amplitude of such an imperfection is set to $0.005\,R_\text{o}$.

The method is implemented in Python through the open source computing platform FEniCS \cite{logg2012automated}. As a linear algebra back--end we used PETSc \cite{balay2017petsc}, the linear Newton iteration is solved in parallel through MUMPS \cite{amestoy2000multifrontal}.

\subsection{Numerical results}
\begin{figure}[t!]
\centering
\includegraphics[width=0.3\textwidth]{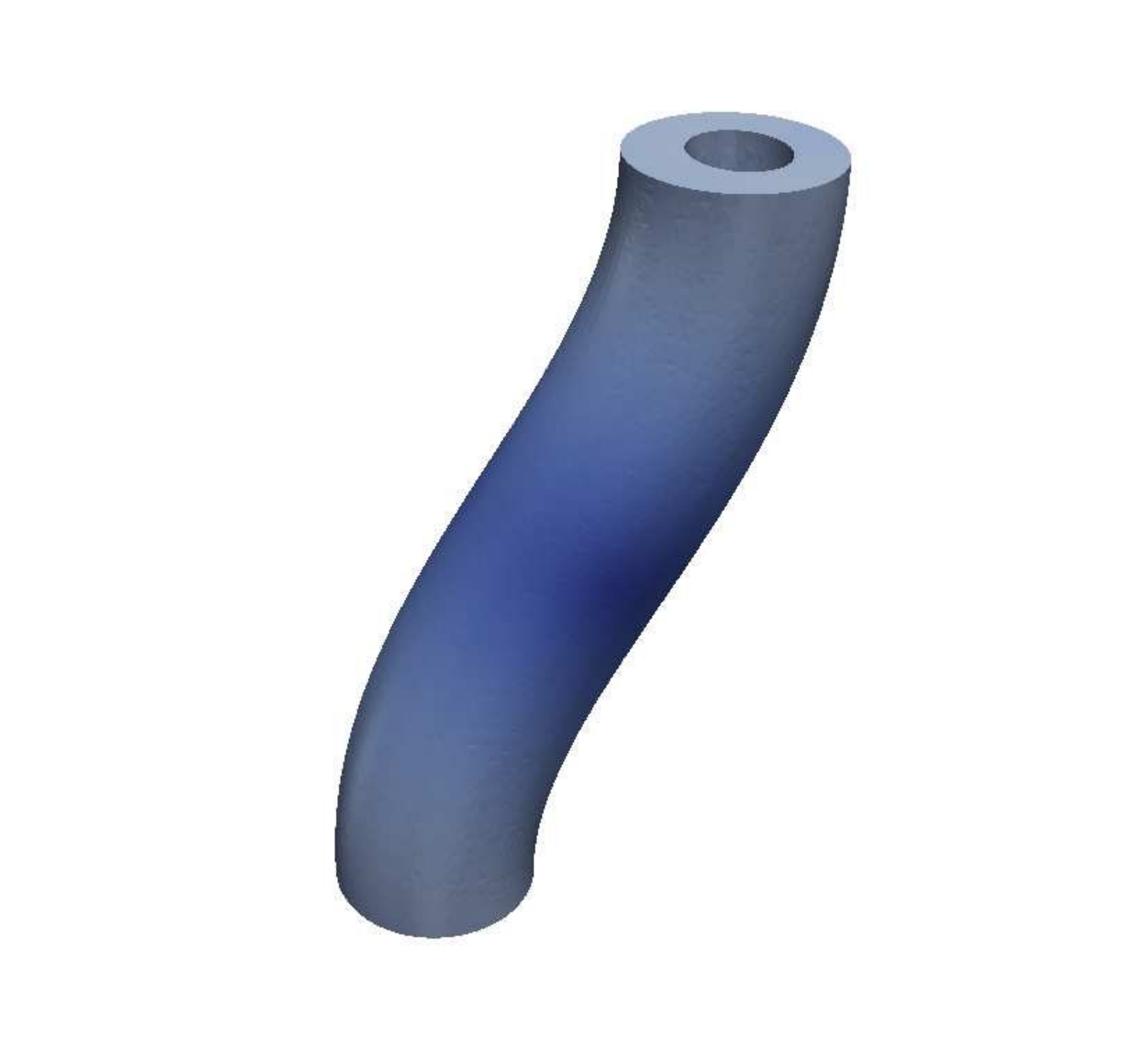}\includegraphics[width=0.3\textwidth]{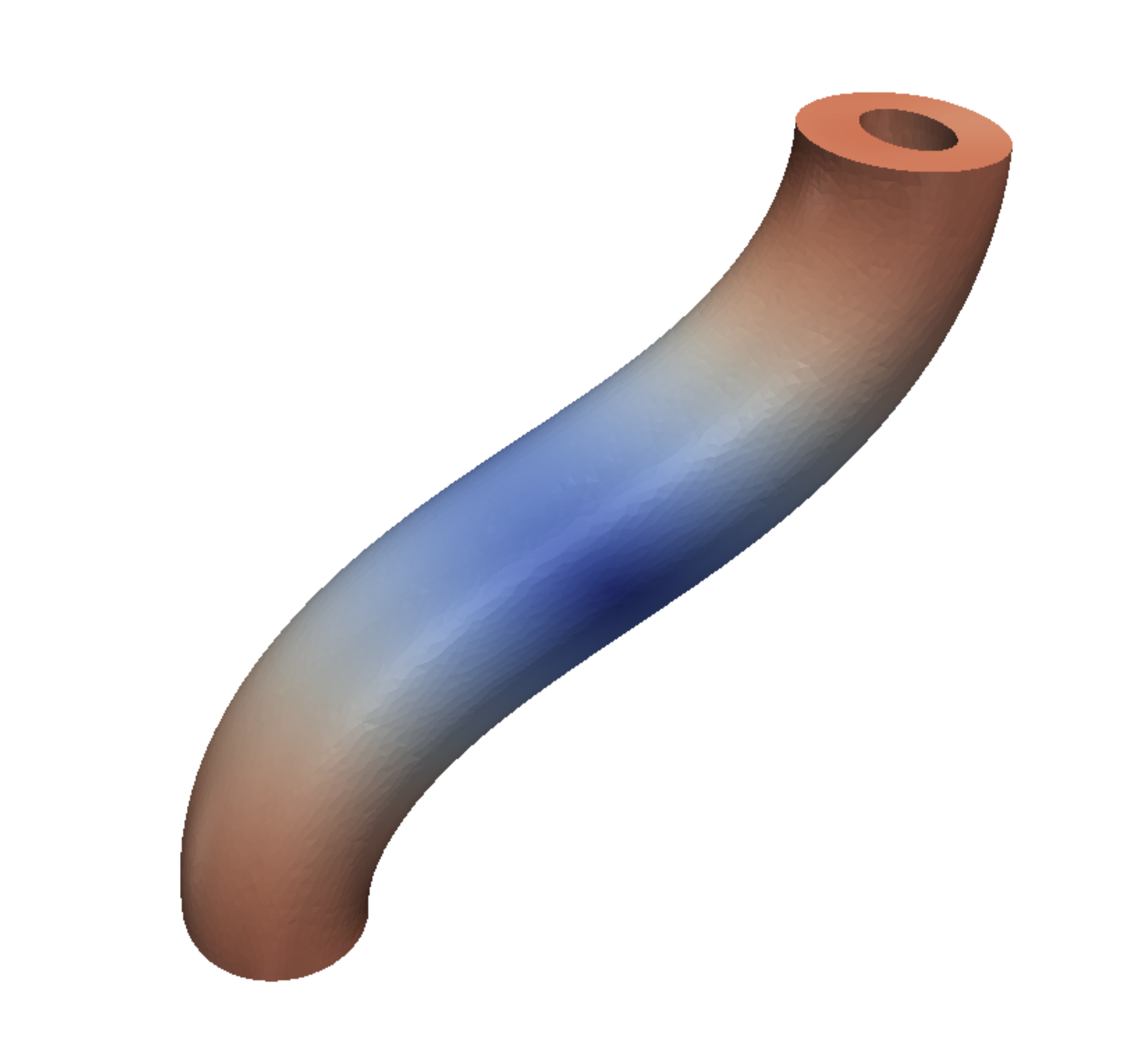}\includegraphics[width=0.3\textwidth]{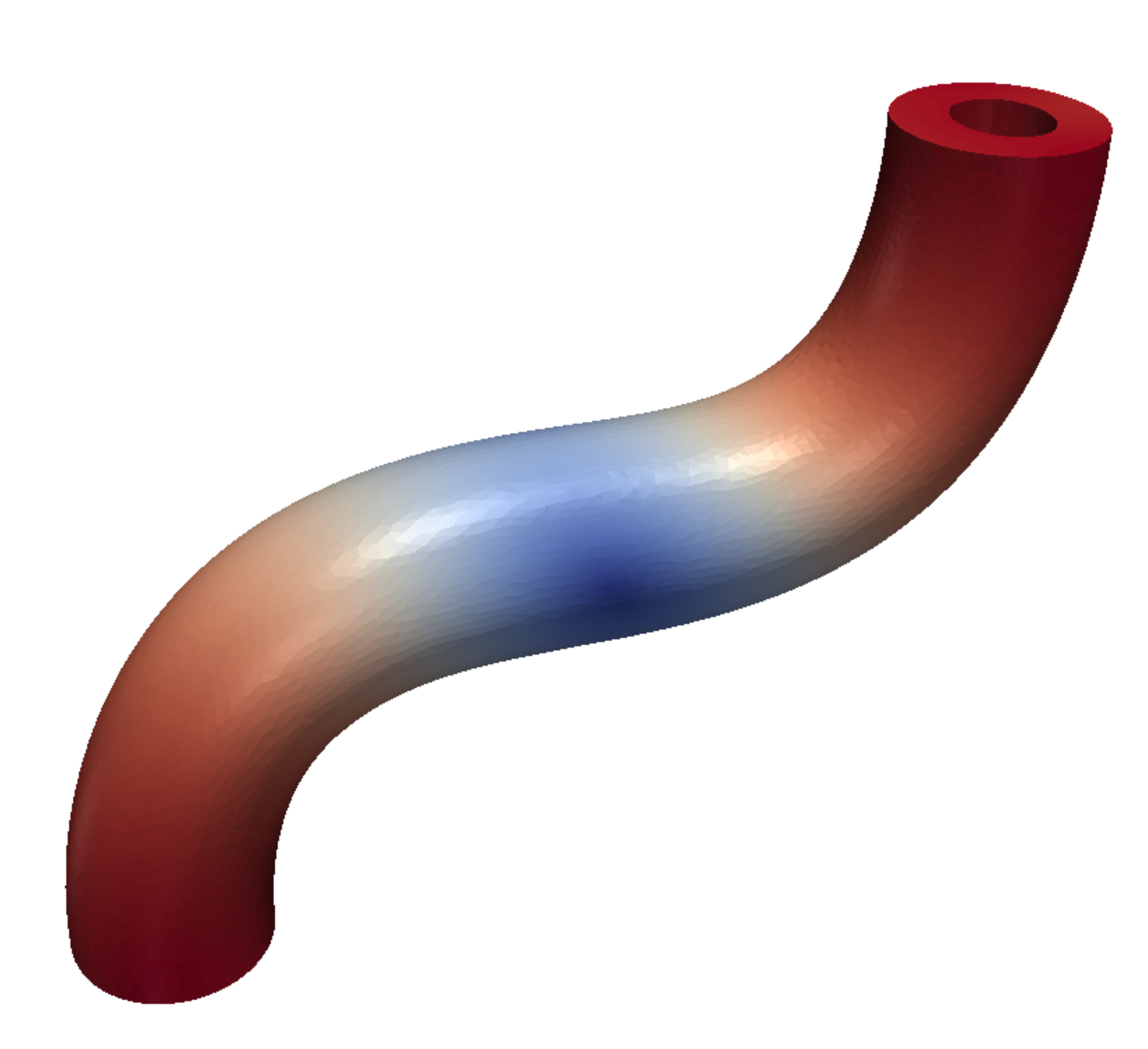}\quad \includegraphics[width=0.07\textwidth]{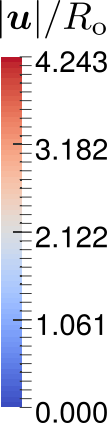}
\caption{Actual configuration of the buckled tube for $\gamma = 1.2$ (left), $\gamma = 1.5$ (center), $\gamma = 1.89$ (right) when $\alpha_k=0.01$, $\alpha_R=0.5$. In such conditions $\gamma_\text{cr} = 1.1103.$}
\label{fig:act_g_full}
\end{figure}
\begin{figure}[t!]
\centering
\includegraphics[width=0.5\textwidth]{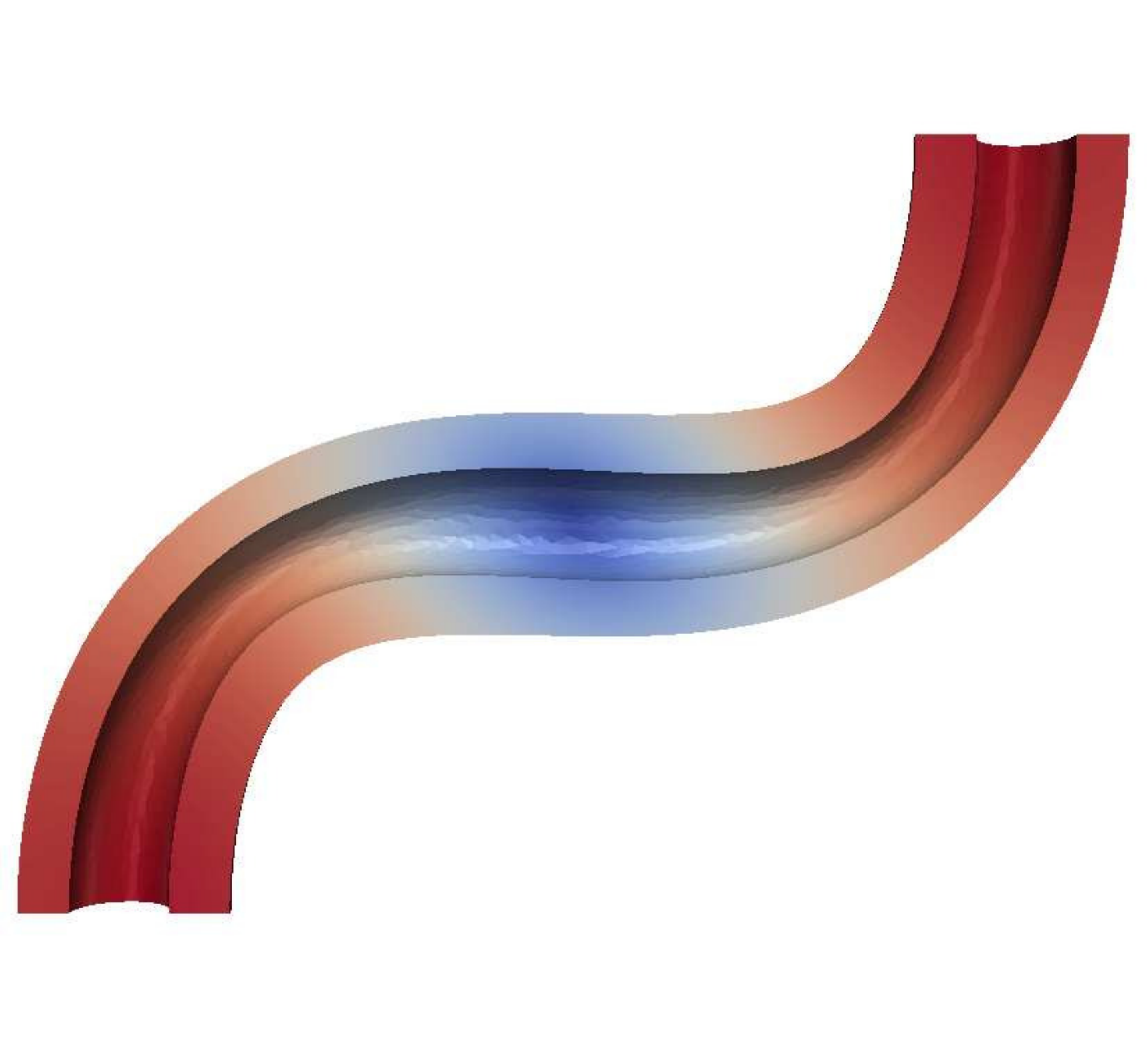}\quad \includegraphics[width=0.07\textwidth]{{leg}.png}
\caption{Actual configuration of the buckled tube for $\gamma = 1.89$. We can notice that the lumen is minimum where the tube has maximum curvature.}
\label{fig:act_g_half}
\end{figure}

In this section we show the results of the numerical simulations for $\alpha_R=0.5$ and $\alpha_k=0.01$. In this case, the critical mode is given by $\tilde{k}_\text{cr}=0.44$, $m_\text{cr}=1$ and the critical threshold is $\gamma_\text{cr} =1.1076$.

We define $\Delta l$ as the average integral of the displacement along the direction $\vect{E}_X$ on the top surface $\mathcal{S}$, namely
\[
\Delta l = \frac{\int_{\mathcal{S}}u_X\,dS}{\pi (R_\text{o}^2-R_\text{i}^2)}.
\]
Such a quantity represents a measure of the displacement of the top surface of the half-cylinder in the direction orthogonal to the axis of the cylinder and parallel to the plane $Y=0$. Since we broke the axial symmetry of the problem by considering an half cylinder only, this is the only plane of symmetry.

In Fig.~\ref{fig:bif} we plot $\Delta l/R_\text{o}$ versus $\gamma$. The numerical results are in agreement with the numerical outcomes, this bifurcation diagram highlights the presence of a supercritical pitchfork bifurcation.

We show the actual configuration of the elastic tube for several values of the control parameter $\gamma$ in Fig.~\ref{fig:act_g_full}. In all the cases, there is a thinning of the tube and a reduction of the lumen in the regions where the curvature is higher as shown in Fig.~\ref{fig:act_g_half}.

The numerical method does not converge near the theoretical marginal stability threshold if $\alpha_k$ is large, probably because the bifurcation becomes subcritical. The improvement of the numerical algorithm is beyond the scope of this article; future works will aim at implementing an arclength continuation method which can also capture the behavior of subcritical bifurcations.

\begin{figure}
\centering
\includegraphics[width=0.25\textwidth]{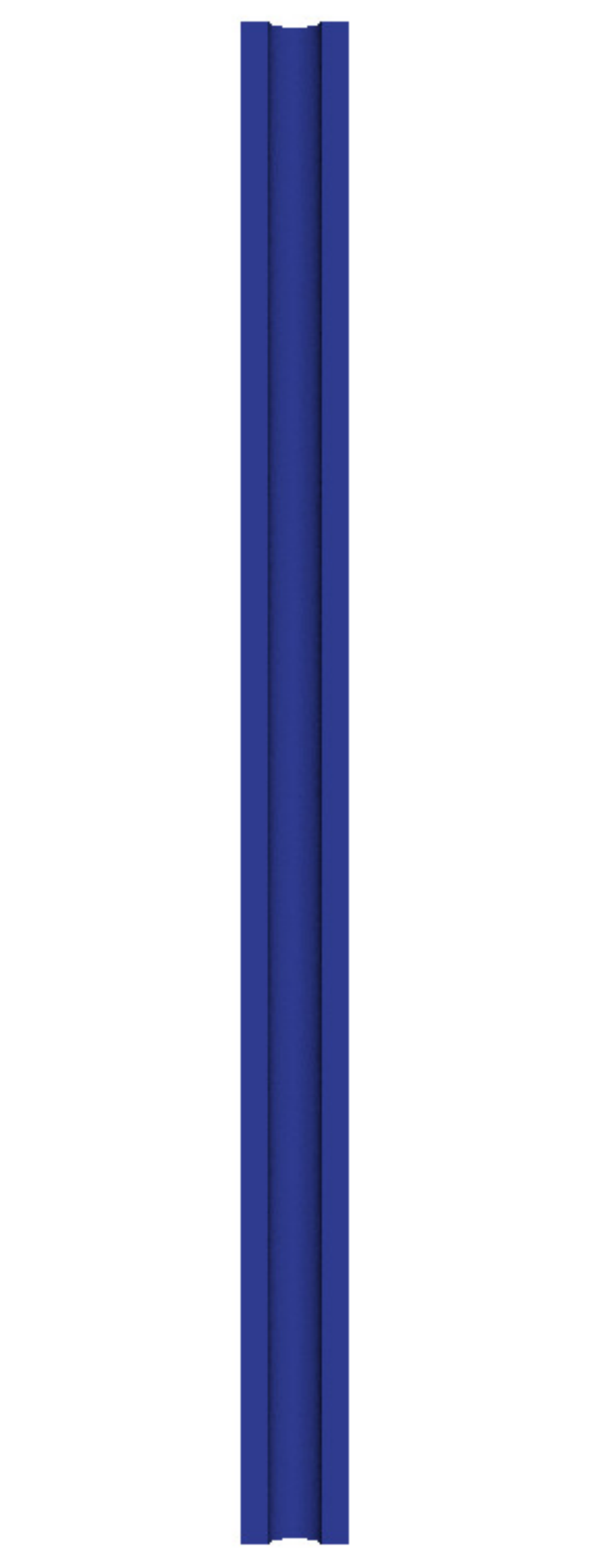}\includegraphics[width=0.25\textwidth]{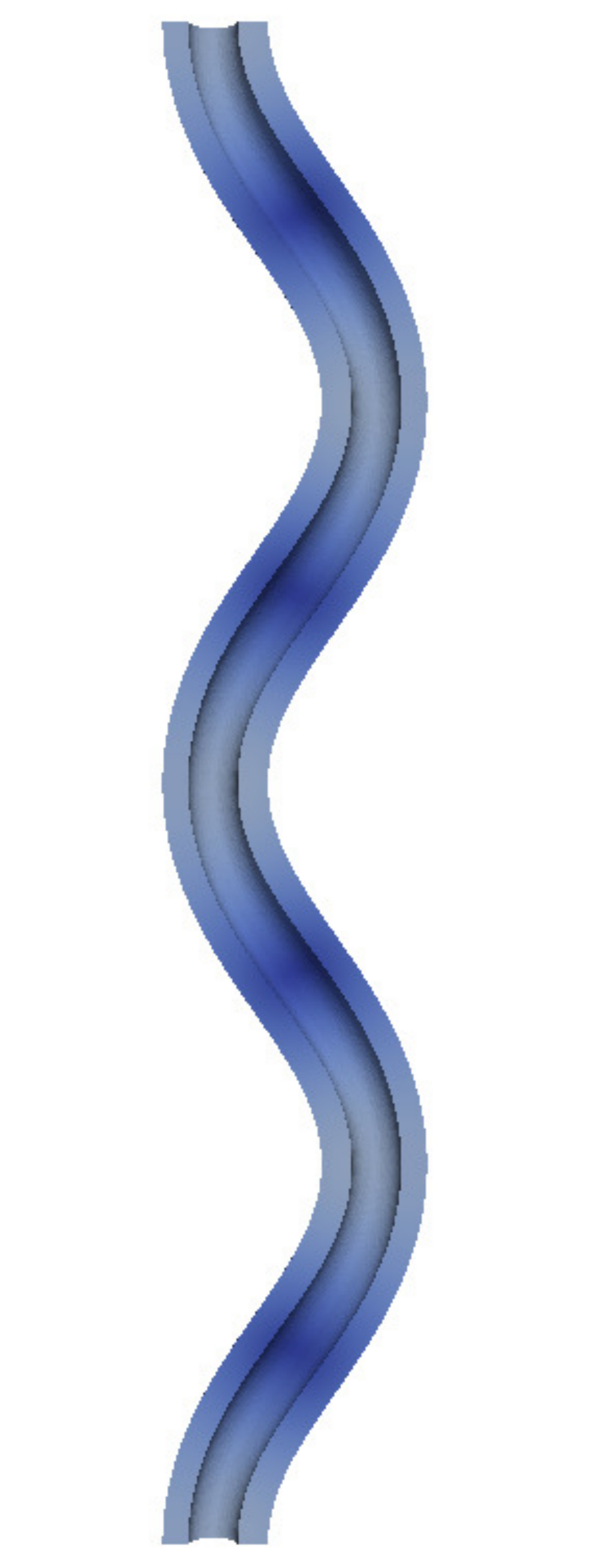}\includegraphics[width=0.25\textwidth]{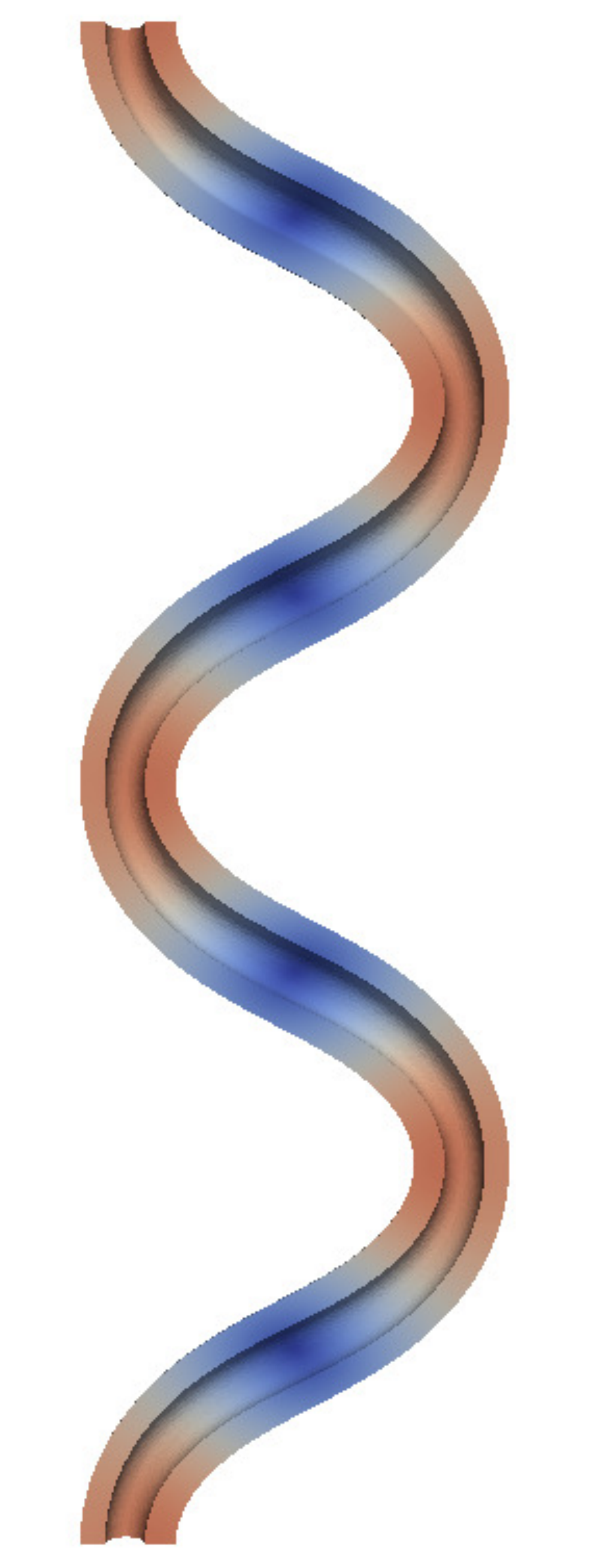}\includegraphics[width=0.25\textwidth]{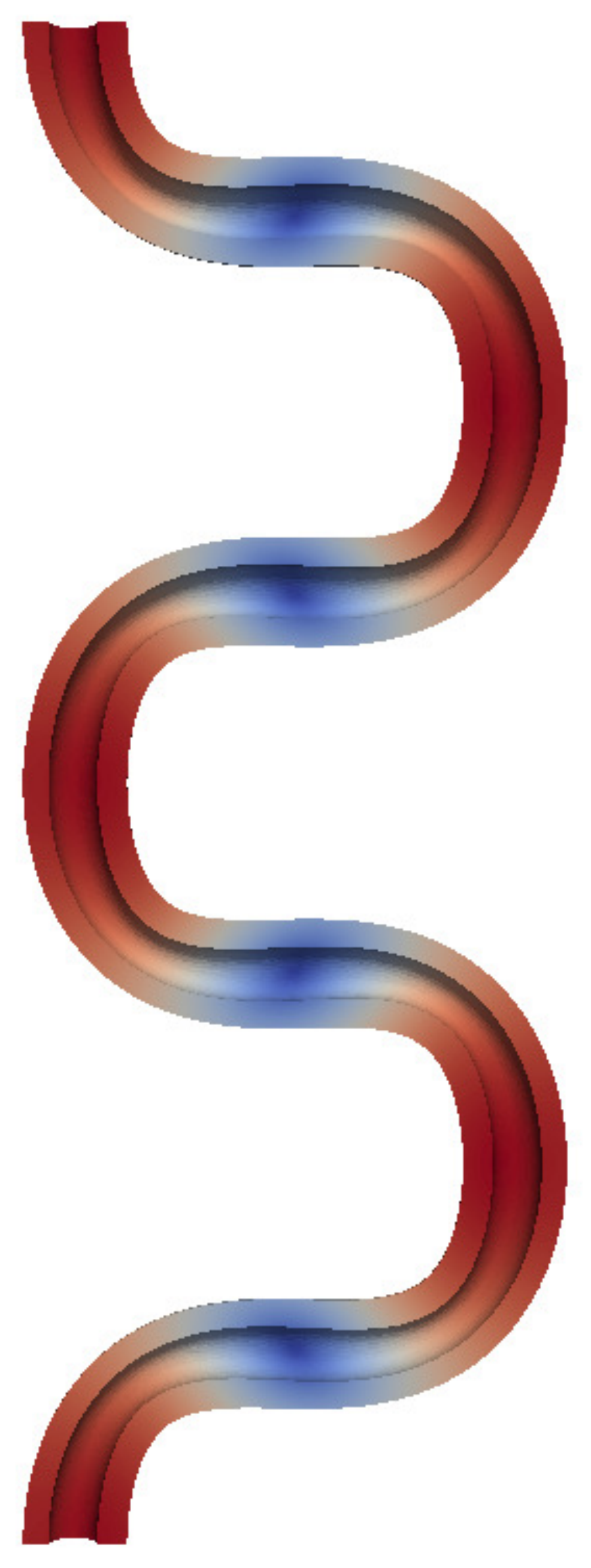}
\caption{Deformed configuration when $\alpha_k = 0.01$, $H/R_\text{o} = 4\pi/\tilde{k}_\text{cr} \simeq 28.55$ and $\gamma = 1,\,1.2,\,1.5,\,1.89$. The corresponding marginal stability threshold is given by $\gamma_\text{cr} = 1.1103$.}
\label{fig:capi}
\end{figure}

If we consider a tube of length $4 \pi /\tilde{k}_\text{cr}\simeq 28.55$ with $\alpha_k=0.01$  we obtain a slenderness ratio which is compatible with the experimental measurements \cite{less1991microvascular}. In Fig.~\ref{fig:capi} we plot the evolution of the tortuosity of the capillary, we observe again that the lumen is minimum in regions where the curvature of the cylinder wall is maximum.

\section{Discussion and concluding remarks}
\label{sec:disc}

In this work, we have proposed a morpho--elastic model of the tortuous shape of tumour vessels. 

In Section \ref{sec:elastic_model}, we have assumed that the tumour capillary is composed of a incompressible neo--Hookean material, and it behaves as  a growing hyperelastic tube that is spatially constrained by a linear elastic environment, representing the surrounding interstitial matter.  We have modeled the growth by using the multiplicative decomposition of the deformation gradient, assuming an incompatible growth along the axial direction, due to the spatial confinement applied at both ends. 

In Section \ref{sec:lin_stab}, we have derived a linear stability analysis on the basic axis--symmetric solution using  the method of incremental deformations superposed on finite strains \cite{ogden1997non}. In order to build a robust numerical method, we exploited the Stroh formulation and the impedance matrix method in order to reduce the incremental boundary value problem to a differential Riccati equation \eqref{eq:Riccati}.

The control parameter of the bifurcation is the axial growth rate $\gamma$, whose critical value is governed by two dimensionless parameters $\alpha_R$ and $\alpha_k$, representing the geometrical aspect ratio  and  the ratio between the energy exerted by the surrounding matter and the bulk strain energy of the capillary, respectively.

The results of the linear stability analysis are collected in Figures \ref{fig:marg_stab_kel_0_RiRo}--\ref{fig:tau}.
Slender capillaries are found having a lower threshold of marginal stability.
On the contrary, when increasing  $\alpha_k$  we find that the overall axial traction load exerted at the tube ends also increases, finding the Euler buckling as the limiting behavior for $\alpha_k\rightarrow 0$ \cite{goriely2008nonlinear}. Interestingly, we find that the linear elastic constraint of the surrounding matter favours the occurrence of short-wavelength critical modes, thus explaining the tortousity of the observed tumour vessels.

The post-buckling behavior is studied implementing numerical simulation using  a mixed finite-element method. The numerical algorithm is based on a Newton based continuation method with an adaptive increment of the control parameter. We considered a physiological geometry for a tumour capillary from referenced literature. The corresponding numerical simulation is validated against the linear stability threshold,  showing that the bifurcation is  supercritical, as depicted in Fig.~\ref{fig:bif}. The emerging morphology of the buckled vessel  is illustrated in  Figs.~\ref{fig:act_g_full}--\ref{fig:capi}. The tortuousity of the capillary is also characterized by lumen restrictions in the localised regions where the capillary reaches its  maximum curvature. This suggests that the elastic bifurcation triggers a significant change in the flow properties inside the vessel. 

In summary, the results of this work show that the tortuosity of the tumour vascular network is mainly driven by the elastic confinement of the interstitial matter where it is embedded. The emerging short-wavelength buckling is similar as the one observed for micro-tubules immersed in the cytosol \cite{brangwynne2006microtubules} and for a growing solid cylinder surrounded by an inert elastic tube \cite{o2013growth}. We remark that modelling  the interstitial matter with linear springs  is a large simplification. Thus, future developments will focus on taking into account for the nonlinear response of the tumour interstitium, possibly including the presence of residual stresses. Moreover, we will investigate the nonlinear effects of the simultaneous buckling of several neighboring capillaries in different spatial networks. The continuation method in numerical simulations shall also be improved  in order to compute the full bifurcation diagram. \cite{seydel2009practical, farrell2016computation}. 
Finally, introducing an elastic-fluid coupling will allow us to quantify how the capillary tortuosity influences the inner fluid transport,  possibly leading to new insights for optimizing  drug delivery withing solid tumors \cite{baish2011scaling, cattaneo2014computational}.

\bibliographystyle{abbrv}
\bibliography{refs}

\appendix
\section{Expressions of the components of the Stroh matrix}
The expression of the sub-block $\tens{N}_1$ of the Stroh matrix is given by
\[
\tens{N}_1=\begin{bmatrix}
 -1 & -m & -k r \\
 \frac{1}{2} m \nu_1 & \frac{-r^2+2 r'^2 R^2+r_\text{i}^2}{2 r'^2 R^2}+\log (r)-\log (r_\text{i}) & 0 \\
 \frac{1}{2} k r \nu_1 & 0 & 0 \\
\end{bmatrix}
\]
where
\[
\nu_1 = \left(\frac{-r^2+2 r'^2 R^2+r_\text{i}^2}{r'^2 R^2}+2 \log (r)-2 \log (r_\text{i})\right).
\]
The sub-block $\tens{N}_2$ reads
\[
\tens{N}_2 = \begin{bmatrix}
 0 & 0 & 0 \\
 0 & \frac{1}{r'^2 \mu } & 0 \\
 0 & 0 & \frac{1}{r'^2 \mu } \\
\end{bmatrix}.
\]
Finally the sub-block $\tens{N}_3$ is given by
\[
\tens{N}_3=\begin{bmatrix}
 N_{41} &N_{42} & N_{43}\\
N_{42} & N_{52} & N_{53}\\
 N_{43} & N_{53} & N_{63} \\
\end{bmatrix}
\]
where
\begin{small}
\begin{align*}
N_{41}=&\frac{k^2 \mu  r^2}{\gamma ^2}-\frac{\left(\left(r^4-2 \left(4 r'^2 R^2+r_\text{i}^2\right) r^2+\left(2 r'^2 R^2+r_\text{i}^2\right)^2\right) m^2\right) \mu }{4 r'^2 R^4}+\\
&+\frac{\left(k^2 r^2 \left(-r^2+2 r'^2 R^2+r_\text{i}^2\right)^2-4 r'^2 R^2 \left(3 r'^2 R^2+r_\text{i}^2\right)\right) \mu }{4 r'^2 R^4}+\\
&+\frac{\mu  \left(2 r'^2 R^2-r'^2 \left(m^2+k^2 r^2\right) (\log (r)-\log (r_\text{i})) R^2\right) (\log (r)-\log (r_\text{i}))}{R^2}+\\
&+\frac{\mu  \left(\left(m^2+k^2 r^2\right) \left(r^2-2 r'^2 R^2-r_\text{i}^2\right)\right) (\log (r)-\log (r_\text{i}))}{R^2}
\end{align*}
\begin{align*}
N_{42}=&\frac{m \mu  \left(-r^4+2 \left(4 r'^2 R^2+r_\text{i}^2\right) r^2+8 r'^4 R^4-r_\text{i}^4\right)}{4 r'^2 R^4}+ \\
&+
\frac{m \mu  \left(4 r'^2 R^2 (\log (r)-\log (r_\text{i})) \left(r'^2 (\log (r_\text{i})-\log (r)) R^2+(r-r_\text{i}) (r+r_\text{i})\right)\right)}{4 r'^2 R^4} 
\end{align*}
\begin{align*}
N_{43}=\frac{k r \mu  \left(-r^2+4 r'^2 R^2+r_\text{i}^2+2 r'^2 R^2 (\log (r)-\log (r_\text{i}))\right)}{2 R^2} 
\end{align*}
\begin{align*}
N_{52}=&\left(3 m^2-1\right) \mu  r'^2+\frac{\left(2 r^2+\left(m^2-1\right) r_\text{i}^2\right) \mu }{R^2}+\frac{k^2 r^2 \mu }{\gamma ^2}+\\
&+\frac{\mu  (\log (r)-\log (r_\text{i})) \left(r^2+2 r'^2 \left(m^2-1\right) R^2-r_\text{i}^2+r'^2 R^2 (\log (r_\text{i})-\log (r))\right)}{R^2}+\\
&-\frac{\left(r^2-r_\text{i}^2\right)^2 \mu }{4 r'^2 R^4}
\end{align*}
\begin{align*}
N_{53}=\frac{k m r \mu  \left(-r^2+3 r'^2 R^2+r_\text{i}^2+2 r'^2 R^2 (\log (r)-\log (r_\text{i}))\right)}{R^2} 
\end{align*}
\begin{align*}
N_{63}=&\frac{r^2 \mu  \left(\left(R^2+\left(-r^2+3 r'^2 R^2+r_\text{i}^2\right) \gamma ^2\right) k^2\right)}{R^2 \gamma ^2}+\\
&+\frac{r^2 \mu  \left(2 r'^2 R^2 \gamma ^2 (\log (r)-\log (r_\text{i})) k^2+m^2 \gamma ^2\right)}{R^2 \gamma ^2}
\end{align*}
\end{small}
\end{document}

%% file: defs.tex
\theoremstyle{plain} 
\DeclareMathOperator{\diver}{div}
\DeclareMathOperator{\grad}{grad}
\DeclareMathOperator{\Diver}{Div}
\DeclareMathOperator{\Grad}{Grad}

\DeclareMathOperator{\tr}{tr}

\DeclareMathOperator{\diag}{diag}

\usepackage{mathtools}
\usepackage{pgfplots}
\pgfplotsset{/pgf/number format/use comma,compat=newest}

\newcommand{\R}{\mathbb{R}}

\newcommand{\vect}[1]{\boldsymbol{#1}}
\newcommand{\tens}[1]{\mathsf{#1}}